\documentclass[a4paper,onecolumn,11pt,accepted=2022-09-30]{quantumarticle}
\pdfoutput=1
\usepackage[utf8]{inputenc}
\usepackage[english]{babel}
\usepackage[T1]{fontenc}
\usepackage{tikz}
\usepackage{lipsum}

\usepackage{arxiv}

\begin{document}
% **************

\title{Model predictive control for robust quantum state preparation}

\author{Andy J. Goldschmidt}
\email{andygold@uw.edu}
\affiliation{Department of Physics, University of Washington, Seattle, WA 98195}
\orcid{0000-0002-2616-9715}

\author{Jonathan L. DuBois}
\affiliation{Lawrence Livermore National Laboratory, Livermore, CA 94550}
\orcid{0000-0003-3154-4273}

\author{Steven L. Brunton}
\affiliation{Department of Mechanical Engineering, University of Washington, Seattle, WA 98195}
\orcid{0000-0002-6565-5118}

\author{J. Nathan Kutz}
\affiliation{Department of Applied Mathematics, University of Washington, Seattle, WA 98195}
\orcid{0000-0002-6004-2275}

\maketitle

\begin{abstract}
A critical engineering challenge in quantum technology is the accurate control of quantum dynamics.
Model-based methods for optimal control have been shown to be highly effective when theory and experiment closely match. 
Consequently, realizing high-fidelity quantum processes with model-based control requires careful device characterization.
In quantum processors based on cold atoms, the Hamiltonian can be well-characterized. For superconducting qubits operating at millikelvin temperatures, the Hamiltonian is not as well-characterized. 
Unaccounted for physics (i.e., mode discrepancy), coherent disturbances, and increased noise compromise traditional model-based control.
This work introduces {\em model predictive control} (MPC) for quantum control applications.
MPC is a closed-loop optimization framework that (i) inherits a natural degree of disturbance rejection by incorporating measurement feedback, (ii) utilizes finite-horizon model-based optimizations to control complex multi-input, multi-output dynamical systems under state and input constraints, and (iii) is flexible enough to develop synergistically alongside other modern control strategies.
We show how MPC can be used to generate practical, optimized control sequences in representative examples of quantum state preparation. 
Specifically, we demonstrate for a qubit, a weakly-anharmonic qubit, and a system undergoing crosstalk, that MPC can realize successful model-based control even when the model is inadequate.
These examples showcase why MPC is an important addition to the quantum engineering control suite.
\end{abstract}

\section{Introduction} \label{sec:introduction}
% ====================
%
Quantum computation can be viewed as an assembly of analogue control pulses driving quantum states toward desired targets~\cite{mckay2018qiskit,wu2020high,ball2021software,li2022pulse,silverio2022pulser}. 
An accurate dynamical model is important for designing optimal control pulses sufficient for each application~\cite{dalessandro2021introduction}. 
In particular, open-loop control strategies using model-based numerical methods have proven to be highly effective in some practical settings, such as for processors based on cold atoms, where models are well-known~\cite{glaser2015training}. 
Sufficiently accurate models are obtained through careful device characterization~\cite{eisert2020quantum}. 
However, descriptions of superconducting qubits~\cite{krantz2019quantum} can fall short due to unknown Hamiltonian terms~\cite{magesan2020effective}, instrument noise, control transfer functions~\cite{hincks2015controlling}, and/or unwanted coupling to unmodelled modes~\cite{werninghaus2021leakage} or other systems~\cite{sarovar2020detecting}. To address these challenges, new frameworks must incorporate measurement feedback in addition to device characterization to optimize pulses for the desired targets. For example, an insufficient pulse from an inaccurate model can be updated retroactively using a model-free optimization derived from experiment feedback~\cite{egger2014adaptive,kelly2014optimal}. 
In this work, we introduce the model predictive control (MPC)~\cite{mayne2000constrained,rakovic2018handbook} framework since it possess a number of structural advantages for closed-loop control of systems with imperfectly known dynamics: (i) MPC inherits a natural degree of disturbance rejection due to measurement feedback, (ii) receding finite-horizon model predictions allow MPC to control complex multi-input multi-output systems under state and control constraints, and (iii) MPC is flexible enough to develop synergistically alongside control innovations like reinforcement learning~\cite{zhang2016learning,gorges2017relations}.  We detail the success of MPC through a number of practical state preparations motivated by superconducting qubit applications.

\begin{figure}
    \centering
    \includegraphics[width=\textwidth]{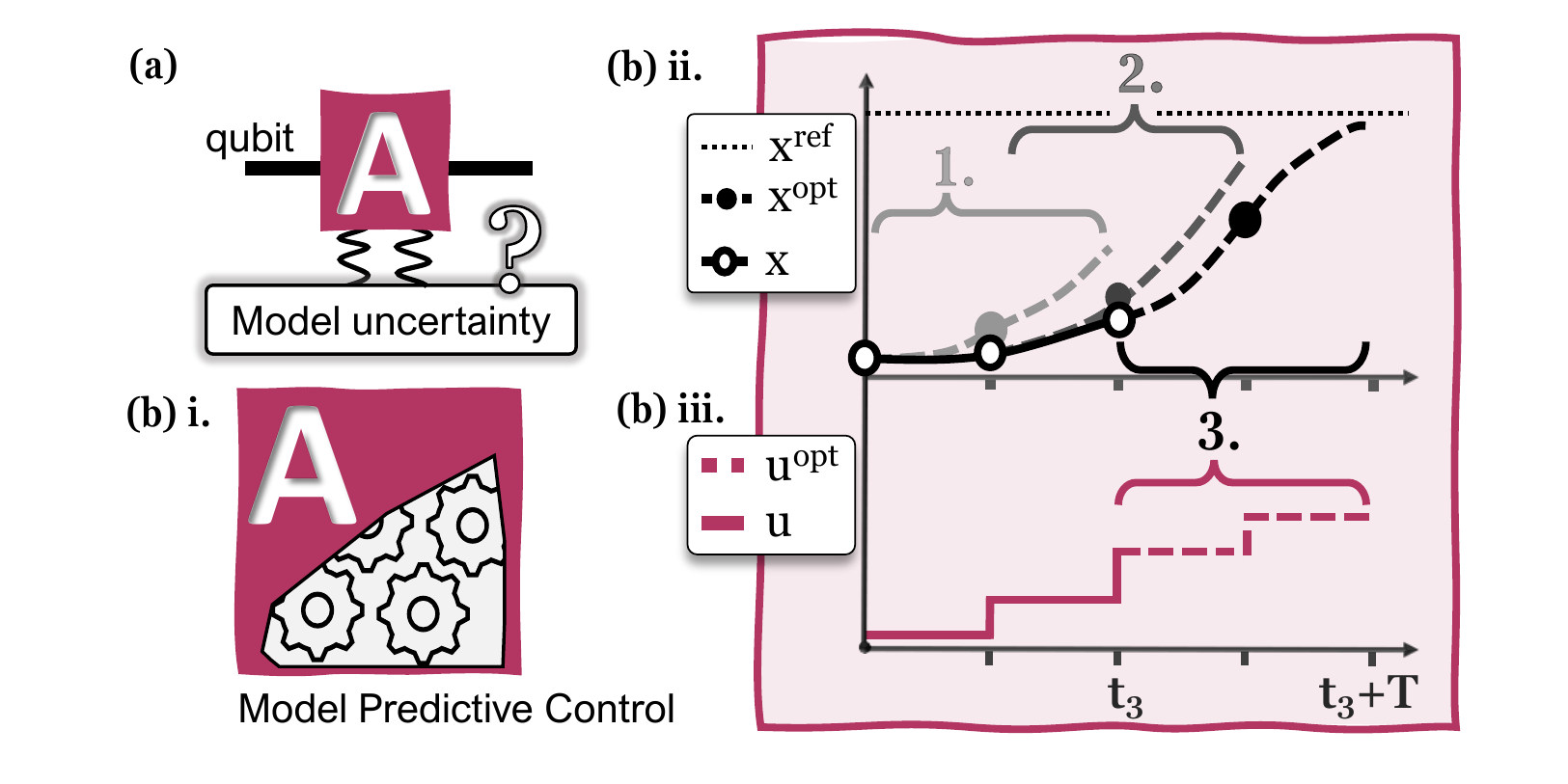}
    \caption{
    \textbf{(a)} In the circuit model, a qubit is represented by a line segment, and a box is draw to indicate a quantum process applied to the qubit. Here, we represent a quantum state preparation abstractly by $\mathbf{A}$ and posit uncharacterized modifications that make $\mathbf{A}$'s model unreliable for open-loop control. \textbf{(b)} Under the hood of the $\mathbf{A}$ operation, robust control pulses are designed using model predictive control~(MPC). In \textbf{(b)~ii-iii}, MPC synthesizes a robust control input by incorporating state feedback into a sequence of receding-horizon control laws (in \textbf{(b)~ii.}, the first three MPC iterations are labeled $1.$, $2.$, and $3.$ and colored with increasing grayscale value). At each MPC iteration, an open-loop control is solved over the current prediction horizon $T$ using $\mathbf{A}$'s unreliable system model to yield $\mathbf{x}^\textrm{opt}(t)$ and $\mathbf{u}^\textrm{opt}(t)$. The first entry of the open-loop solution $\mathbf{u}^\textrm{opt}(t)$  is applied as $\mathbf{u}(t)$ in \textbf{(b)~iii.}, and the resulting state $\mathbf{x}(t+1)$ is recorded in \textbf{(b)~ii.} The next MPC iteration begins from this recorded state.
    }
    \label{fig:main}
\end{figure}

MPC~\cite{mayne2000constrained,rakovic2018handbook} (see Figure~\ref{fig:main}) is a closed-loop optimization strategy that compliments existing approaches -- that is, MPC is not a new type of controller on its own. Instead, standard MPC implements a sequence of open-loop controllers \textit{online} as a function of the current state. First, a model prediction is made out to some finite horizon and used to inform the present control decision. Second, the decision is implemented, and the prediction begins again from the new measured state. For this reason, MPC is often referred to as a receding-horizon strategy. MPC has a history of practical successes ranging from the chemical process industry~\cite{qin2003survey} to autonomous vehicles~\cite{falcone2007predictive} and reusable rocket landings~\cite{eren2017model}. 
Notably, MPC for quantum dynamics is closely connected to MPC for classical control using Koopman-von Neumann theory, a description of classical mechanics using a Hilbert space of observables defined on phase-space. Koopman-von Neumann theory shares strong historical connections with quantum mechanics and has wide applicability to modern data-driven analysis~\cite{koopman1931hamiltonian,koopman1932dynamical,vonneumann1932mathematical,Mezic2005nd,Mezic2013arfm,brunton2021modern}. Connections between Koopman-von Neumann theory and quantum control dynamics were studied in Reference~\cite{goldschmidt2021bilinear}. Recent successes using MPC within the classical Koopman-von Neumann theory~\cite{abraham2017model,korda2018linear,peitz2019koopman,peitz2020data,bruder2021advantages,folkestad2021koopman} can motivate synergistic development of MPC for quantum dynamics. In this work, we consider planning for control using fixed model discrepancies within MPC. In practice, model discrepancies can be improved by integrating data-driven modelling together with MPC. The integration of planning over a model and learning of a model requires making choices about how to best utilize the available data. There are many trade-offs to consider such as accuracy, interpretability, and data efficiency: for a review, see Reference~\cite{moerland2020model}. One approach to the integration of planning and learning is data-driven model synthesis based on Koopman theory, using streaming data~\cite{pendergrass2016streaming,zhang2019online,giannakis2021learning} to build or improve operator-theoretic models of the dynamics in an \textit{online} way as the MPC horizon recedes. In the language of Reference~\cite{moerland2020model}, this approach is known as \textit{planning over a learned model}, and is exemplified by approaches like Embed to Control (E2C)~\cite{watter2015embed}. Error bounds for Koopman models of control systems were recently studied in \cite{nuske2021finite,schaller2022guaranteed}, enabling the rigorous analysis of data-driven MPC schemes.

There is a distinction to make between the classical and quantum settings. When MPC is used in robotics, the implementation of a control decision results in the actual motion of the robot to a new position. Unlike the robot, a quantum state is a statistical outcome from the measurement of many identically-prepared quantum experiments. After running successive or parallel experiments to realize quantum state tomography~\cite{eisert2020quantum}, the next quantum experiment is reset at the initial (or perhaps projected) state. There is a possibility of using the state information to look backward and improve upon the previous model or pulse. Backward-looking, model-free algorithms iteratively search for parameter changes that improve upon desired control objectives. Such methods have been highly successful in calibrating controls to quantum experiments~\cite{egger2014adaptive,kelly2014optimal,wittler2021integrated}. However, many iterations may be necessary to successfully search the parameter space; hence, many evaluations of the control objective may be required. Moreover, free parameters are often limited in number to maintain the feasibility of the optimization. The alternative MPC perspective is to solve the quantum problem as if it is an \textit{online} optimization. The forward-looking MPC trades iterative improvements for extra online control time by accepting and proceeding from the current quantum state tomography outcome (we return to this comparison in detail in Appendix~\ref{apdx:neldermead} after establishing an explicit context via our numerical examples). The resulting MPC control scheme is robust, as MPC inherits a natural degree of disturbance rejection due to the feedback. Moreover, MPC can handle the many parameters of multi-input, multi-output systems. Finally, MPC has an expansive literature base from which to build. In this paper, we demonstrate the utility of the MPC perspective in the context of quantum control engineering. First, we introduce our MPC implementation. Then, we proceed with examples of robust quantum state preparation for ideal qubits, weakly-anharmonic qubits, and qubits coupled by undesired crosstalk.

\section{MPC for Quantum Control}
% ===============================

In what follows, Section~\ref{sec:mpc} describes the theoretical framework for standard MPC, restricted for simplicity to linear models with quadratic cost functions. Section~\ref{sec:quantumdynamics} describes the nonlinear equations of motion for quantum control dynamics, and Section~\ref{sec:nmpc} shows how to modify standard MPC for the nonlinear quantum control dynamics by way of nonlinear MPC.

\subsection{Background: MPC} \label{sec:mpc}
% --------------------------
The standard linear-quadratic MPC involves a discrete-time linear model, a quadratic cost function, and possibly linear constraints on the state and control. The optimization problem generated under these assumptions is a quadratic program (QP). With an initial state $\mathbf{x}_0$ and reference trajectories $\mathbf{X}^{\textrm{ref}} = \left[\mathbf{x}^{\textrm{ref}}(0), \mathbf{x}^{\textrm{ref}}(1), \dots, \mathbf{x}^{\textrm{ref}}(T) \right]$ and $\mathbf{U}^{\textrm{ref}} = \left[\mathbf{u}^{\textrm{ref}}(0), \mathbf{u}^{\textrm{ref}}(1), \dots, \mathbf{u}^{\textrm{ref}}(T-1) \right]$, the QP is given by
\begin{subequations}
\begin{align}
    \textrm{QP}_{\textrm{MPC}}(\mathbf{x}_0, \mathbf{X}^\textrm{ref}, \mathbf{U}^\textrm{ref}) = & \label{eqn:qpmpc} \\
    \arg\min_{\mathbf{X}, \mathbf{U}} \, 
    \sum_{t=0}^{T-1} &\left( \norm{\mathbf{x}(t) {-} \mathbf{x}^\textrm{ref}(t)}^2_\mathbf{Q} {+} \norm{\mathbf{u}(t) {-} \mathbf{u}^\textrm{ref}(t)}^2_\mathbf{R} \right) {+} \norm{\mathbf{x}(T) {-} \mathbf{x}^\textrm{ref}(T)}^2_{\mathbf{Q}_f} \label{eqn:quadraticcost} \\
    \textrm{s.t.}\quad 
    &\mathbf{x}(t+1) = \mathbf{A}(t) \mathbf{x}(t) + \mathbf{B}(t) \mathbf{u}(t), \quad t=0,1,\dots,T-1 \label{eqn:linearconstr} \\
    &\mathbf{x}(0) = \mathbf{x}_0 \\
    &\mathbf{x}_\textrm{min} \le \mathbf{x}(t) \le \mathbf{x}_\textrm{max}, \quad t=1,2,\dots,T \\
    &\mathbf{u}_\textrm{min} \le \mathbf{u}(t) \le \mathbf{u}_\textrm{max}, \quad t=0,1,\dots,T-1
\end{align}
\end{subequations}
where $\norm{\mathbf{x}}^2_\mathbf{Q} := \mathbf{x}\T \mathbf{Q} \mathbf{x}$. The quadratic costs $\mathbf{Q}$, $\mathbf{R}$, $\mathbf{Q}_f$, dynamics $\mathbf{A}(t)$, $\mathbf{B}(t)$, and prediction horizon $T$ are parameters. The QP returns an optimal state and control, $\mathbf{X}^\textrm{opt}$ and $\mathbf{U}^\textrm{opt}$, for the entire prediction horizon. If no state or input constraints are introduced, then this problem could be solved using familiar linear quadratic methods~\cite{anderson2007optimal}. Alternatively, more general constraints may be considered if the resulting optimization can be reformulated as a second-order cone program (SOCP), of which a QP is a special case~\cite{boyd2004convex,jackson2021altro}. Typical MPC applications involve real-time control of complex systems. This motivates work on faster and more efficient MPC algorithms~\cite{wang2009fast}. Many powerful open-source and commercial MPC solvers exist today for a variety of purposes.  In our work, we used CVXPY~\cite{diamond2016cvxpy} (a research-friendly Python-embedded convex-programming library) together with the QP solver OSQP~\cite{stellato2021osqp}.

\subsection{Quantum control dynamics} \label{sec:quantumdynamics}
% -----------------------------------
%
Markovian quantum control dynamics~\cite{altafini2012modelling} are bilinear with respect to the state vector $\ket{\psi(t)} \in \C^N$ and coherent control amplitudes $u_j(t)$ such that
\begin{equation}
\frac{\partial}{\partial t} \ket{\psi(t)} = -i ({H}_0 + \sum_j u_j(t) {H}_j ) \ket{\psi(t)} = {H}(t) \ket{\psi(t)}.
\end{equation}  
For a qubit, $N=2$, while for a qudit, $N=d$. A sequence of $n$ qudits forms a quantum register which can be used for quantum computations; in this case, the state grows exponentially with respect to the length of the register, $N = d^n$. An ensemble of pure quantum states or registers can be completely characterized, in the sense of their measurement statistics, by a density matrix $\rho(t)$; that is, a non-negative self-adjoint operator in $\C^{N \times N}$ with trace one. The density matrix enables a simple characterization of Markovian interactions by the environment on the state in terms of dissipative dynamical operators; this is the Markovian master equation description of an open quantum system, 
\begin{equation} \label{eqn:lindblad}
    \frac{\partial}{\partial t} \rho(t) = -i[H(t), \rho(t)] + \frac{1}{2} \sum_{j,k=1}^{N^2-1} c_{jk} \left( [D_j, \rho(t) D_k^\dagger] + [D_j \rho(t), D_k^\dagger] \right) \,,
\end{equation}
where $H(t)$ is a trace-zero Hermitian operator (corresponding with the system Hamiltonian), $\{D_j\}_{j=1}^{N^2-1}$ is an orthonormal set of complex matrices with trace zero, and $C:=(c_{jk})$ is positive semi-definite. For a closed quantum system $C = 0$, the equation simplifies to the quantum Liouville equation. In this paper, we use the density matrix to represent the quantum state for the purpose of describing control experiments for quantum state preparation. In order to continue to represent our density matrix as a state vector in line with our MPC intentions, we apply the vectorization operation (see Appendix~\ref{apdx:vec}) 
\begin{equation}
    \begin{pmatrix}
        \rho_{00} & \rho_{01} \\
        \rho_{10} & \rho_{11}
    \end{pmatrix} \overset{\textrm{vec}}{\mapsto}
    \begin{pmatrix} \rho_{00} \\ \rho_{01} \\ \rho_{10} \\ \rho_{11} \end{pmatrix} .
\end{equation}
In addition, where it is necessary to work with completely real state vectors we rely on the isomorphism between complex numbers and $2 \times 2$ matrices for the operators,
\begin{equation}
    a + b i \overset{\textrm{real}}{\mapsto} \begin{pmatrix} a & -b \\ b & a \end{pmatrix}
\end{equation}
To apply MPC for quantum control dynamics, the constraint in Equation~\eqref{eqn:linearconstr} is replaced by the discretization of the bilinear dynamics in Equation~\eqref{eqn:lindblad}~\cite{goldschmidt2021bilinear}. To first order it is given by
\begin{equation} \label{eqn:eulerbilinear}
    \mathbf{x}(t + 1) = (\mathbf{A} + \sum_j u_j(t) \mathbf{N}_j) \mathbf{x}(t) .
\end{equation}

Our intended task in this paper is quantum state preparation, in which a quantum state represented by $\rho$ is driven to realize a reference state $\rho^\textrm{ref}$. The success of quantum state preparation is typically scored using the (squared) fidelity $F(\rho, \rho^\textrm{ref})=\Tr\{\sqrt{\sqrt{\rho} \rho^\textrm{ref} \sqrt{\rho}}\}^2$~\cite{kliesch2021theory}. In the common case where $\rho^\textrm{ref}$ is a pure state, $F(\rho, \rho^\textrm{ref}) = \Tr{\rho \rho^\textrm{ref}}$. Fidelity approximates the distance measures of quantum states given by the Schatten $p$-norms, $\norm{\rho - \rho^\textrm{ref}}_p = \norm{\sigma(\rho - \rho^\textrm{ref})}_p = (\sum_j \sigma_j^p)^{1/p}$ where $\sigma(\cdot)$ returns the vector of singular values $\sigma_j$ of a given matrix. In Equation~\eqref{eqn:quadraticcost}, we have formulated the MPC objective in the standard way as a least-squares cost, which penalizes deviations from a reference trajectory. In the case of the vectorized density matrix, this choice is consistent with overlap fidelity in the following way: if $\mathbf{x}$ is the corresponding vectorization of the density matrices, their norms are connected under the identity $\norm{\rho}_2 = \norm{\mathbf{x}}_2$ (Appendix~\ref{apdx:opnorm}).
Quantum state preparation can be understood as a subroutine to design the control pulses necessary to realize a reference quantum process. In Appendix~\ref{apdx:gates}, we discuss how MPC for quantum state preparation can be naturally extended to the pursuit of quantum gate synthesis, where the goal is to realize a finite set of unitary processes which can be combined to yield arbitrary quantum computations.

\subsection{Nonlinear MPC} \label{sec:nmpc}
% ------------------------
%
\begin{algorithm}[t]
\caption{Sequential Quadratic Programming}
\label{alg:sqp}
\begin{algorithmic}[1]
		\Statex{\textbf{INPUT: } Initial state $\mathbf{x}_0$, initial guesses $(\mathbf{X}^\textrm{guess}, \mathbf{U}^\textrm{guess})$, and references $(\mathbf{X}^\textrm{ref}, \mathbf{U}^\textrm{ref})$.}
		\Statex{\textbf{OUTPUT: } Optimal trajectories $(\mathbf{X}^\textrm{opt}, \mathbf{U}^\textrm{opt})$.}
		\Function{SQP}{$\mathbf{x}_0$, $\mathbf{X}^\textrm{ref}$, $\mathbf{U}^\textrm{ref}$, $\mathbf{X}^\textrm{guess}$, $\mathbf{U}^\textrm{guess}$}
	    \While{Not converged}
	        \State{Compute the linearizations $\mathbf{A}^\textrm{guess}(t)$, $\mathbf{B}^\textrm{guess}(t)$}
	        \Comment{Eq.~\eqref{eqn:linearization}}
	        \State{$(\mathbf{X}^\textrm{opt}, \mathbf{U}^\textrm{opt}) \leftarrow \textrm{QP}_\textrm{NMPC}(\mathbf{x}_0, \mathbf{X}^\textrm{ref}, \mathbf{U}^\textrm{ref}, \mathbf{X}^\textrm{guess}, \mathbf{U}^\textrm{guess})$}
	        \Comment{Eq.~\eqref{eqn:nonlinearconstr}}
	        \State{Compute $\alpha \in [0, 1]$ using line search}
	        \State{$(\mathbf{X}^\textrm{guess}, \mathbf{U}^\textrm{guess}) \leftarrow (\mathbf{X}^\textrm{guess}, \mathbf{U}^\textrm{guess}) + \alpha (\mathbf{X}^\textrm{opt} - \mathbf{X}^\textrm{guess}, \mathbf{U}^\textrm{opt} - \mathbf{U}^\textrm{guess})$}
	    \EndWhile
	    \State{$(\mathbf{X}^\textrm{opt}, \mathbf{U}^\textrm{opt}) \leftarrow (\mathbf{X}^\textrm{guess}, \mathbf{U}^\textrm{guess})$} 
    	\EndFunction
\end{algorithmic}
\end{algorithm}

Continuous-time quantum control dynamics are bilinear. By introducing a general nonlinear dynamics constraint $\mathbf{x}(t+1) = \mathbf{f}(\mathbf{x}(t), \mathbf{u}(t), t)$ in place of the linear dynamics found in Equation~\eqref{eqn:linearconstr}, we enter the realm of nonlinear MPC. Here, the necessary optimization problems are typically non-convex. Nonlinear MPC permits the use of more complex models, so applications are abundant. In recent years, practical realizations have improved alongside advances in numerical methods and non-convex optimization~\cite{rakovic2018handbook}. In this paper, we treat the nonlinear MPC by following the sequential quadratic program (SQP) approach to solve the open-loop optimization between MPC steps. The SQP approach is outlined in Algorithm~\ref{alg:sqp}~\cite{gros2020from}. SQP is a type of direct optimization: the state and control are explicitly treated as optimization variables which are constrained by the dynamics~\cite{propson2022robust}. This is in contrast to indirect approaches (in quantum control, familiar examples include GRAPE~\cite{khaneja2005optimal}, GOAT~\cite{machnes2018tunable}, or Krotov’s method~\cite{goerz2019krotov}), where the dynamics are used to eliminate the optimization over the state variables. SQP iteratively solves local QP approximations of the nonlinear optimization problem until convergence is achieved. To account for the approximate character of the problem, steps are taken by following a line search~\cite{nocedal2006numerical}. Each local QP approximation is taken about guess trajectories $\mathbf{X}^\textrm{guess}=\left[\mathbf{x}^{\textrm{guess}}(0), \mathbf{x}^{\textrm{guess}}(1), \dots, \mathbf{x}^{\textrm{guess}}(T) \right]$ and $\mathbf{U}^\textrm{guess}=\left[\mathbf{u}^{\textrm{guess}}(0), \mathbf{u}^{\textrm{guess}}(1), \dots, \mathbf{u}^{\textrm{guess}}(T-1) \right]$). Denote this modified version of the QP as $\textrm{QP}_\textrm{NMPC}(\mathbf{x}_0, \mathbf{X}^\textrm{ref}, \mathbf{U}^\textrm{ref}, \mathbf{X}^\textrm{guess}, \mathbf{U}^\textrm{guess})$. To run this modified optimization, we first linearize the dynamics $\mathbf{f}(\mathbf{x}(t), \mathbf{u}(t), t)$ such that
\begin{align} \label{eqn:linearization}
    & \mathbf{A}^\textrm{guess}(t) = \frac{\partial \mathbf{f}}{\partial \mathbf{x}}\Big|_{\mathbf{x}^\textrm{guess}(t),\, \mathbf{u}^\textrm{guess}(t)} \\
    & \mathbf{B}^\textrm{guess}(t) = \frac{\partial \mathbf{f}}{\partial \mathbf{u}}\Big|_{\mathbf{x}^\textrm{guess}(t),\, \mathbf{u}^\textrm{guess}(t)} \nonumber.
\end{align} 
Then, we replace Equation~\eqref{eqn:linearconstr} in $\textrm{QP}_\textrm{MPC}(\mathbf{x}_0, \mathbf{X}^\textrm{ref}, \mathbf{U}^\textrm{ref})$ with 
\begin{align} \label{eqn:nonlinearconstr}
    &\mathbf{x}(t+1) = \mathbf{x}^\textrm{guess}(t + 1) + \mathbf{A}^\textrm{guess}(t) \Delta \mathbf{x}(t) + \mathbf{B}^\textrm{guess}(t) \Delta \mathbf{u}(t) + \mathbf{r}(t+1) \\
\intertext{where} \nonumber \\
    &\Delta \mathbf{x}(t) := \mathbf{x}(t) - \mathbf{x}^\textrm{guess}(t) \nonumber \\
    &\Delta \mathbf{u}(t) := \mathbf{u}(t) - \mathbf{u}^\textrm{guess}(t) \nonumber \\
    & \mathbf{r}(t+1) := \mathbf{f}(\mathbf{x}^\textrm{guess}(t), \mathbf{u}^\textrm{guess}(t), t) -  \mathbf{x}^\textrm{guess}(t + 1). \nonumber
\end{align} 

The discretization of the continuous time quantum control dynamics can be taken before or after the linearization in Equation~\eqref{eqn:linearization}~\cite{gros2020from}. We pursue the case where discretization is taken first. For quantum control dynamics, we saw in Equation~\eqref{eqn:eulerbilinear} that the discretized model contains a bilinear nonlinearity in the state and control. A bilinear nonlinearity is relatively simple, and it has been shown that successful and efficient nonlinear MPC controllers can be implemented even under guesses that account only for the initial state (i.e. $\mathbf{x}^\textrm{guess}(t) = \mathbf{x}_0$ for $t=0, 1, \dots T-1$)~\cite{ccimen2008state,bruder2021advantages,folkestad2021koopman}. The known bilinear structure can also be used to increase optimization efficiency~\cite{peitz2020data}. Additional accuracy can be sought by using better discretizations or even data-driven numerical integrators~\cite{gros2020from, goldschmidt2021bilinear}. Automatic differentiation tools can also prove impactful in these cases. 

SQP can be warm-started after the implementation of the first control decision. At the initial timestep $t=0$, suppose SQP was run in full to find an optimal state and control over the prediction horizon. As a by-product, good guess trajectories $\mathbf{X}_{0}^\textrm{guess}$ and $\mathbf{U}_{0}^\textrm{guess}$ for this prediction horizon have been obtained through SQP. All future guess trajectories can then be found via a shifting procedure applied to these good guess trajectories. A common practical implementation is as follows: obtain the warm-start guess trajectories for the initial timestep $t$ from the previous $\mathbf{X}_{t-1}^\textrm{guess}$ and $\mathbf{U}_{t-1}^\textrm{guess}$ by eliminating the first value and duplicating the final value so that each sequence retains its length. Heuristically, the shifted guesses are close to the optimal value of the nonlinear program. Therefore, for all initial timesteps $t > 0$ the SQP can be terminated after just one iteration with $\alpha = 1$ (see Algorithm~\ref{alg:sqp})~\cite{gros2020from}.

\section{MPC for robust quantum state preparation}
% ================================================
%
MPC can be utilized for many control strategies, such as  setpoint stabilization (realizing a static reference state), tracking (following a time-dependent reference trajectory), and path following (staying close to any part of a time-independent reference trajectory at all times)~\cite{rakovic2018handbook}. Each strategy has different objectives and criteria for convergence and stability. Selecting a suitable control strategy is important for obtaining the best performance. In our examples, we consider MPC for setpoint stabilization to perform robust quantum state preparation in the presence of coherent noise and modelling inaccuracies.

Section~\ref{sec:qubit} introduces MPC for robust state preparation of a qubit, and discusses the implications of the measurement feedback period. Section~\ref{sec:waqubit} uses MPC to control a weakly anharmonic qubit assuming no model knowledge of the anharmonicity and compares the MPC treatment with an analytic pulse design. Finally, Section~\ref{sec:crosstalk} looks at simultaneous state preparation of coupled qubits in the presence of an unmodelled coupling (i.e. crosstalk). Working in the individual (reduced) qubit spaces, MPC is used to overcome the crosstalk effect. For each of our ground truth simulations, we use the QuTiP Python package~\cite{johansson2012qutip, johansson2013qutip2}. The following examples report the Hamiltonians of each toy system, with the understanding that the relevant modifications are made following Section~\ref{sec:quantumdynamics} to obtain the dynamical model used to constrain the optimal control problem.

\subsection{Qubit} \label{sec:qubit}
% ----------------
%
\begin{figure}
    \centering
    \includegraphics[width=\textwidth]{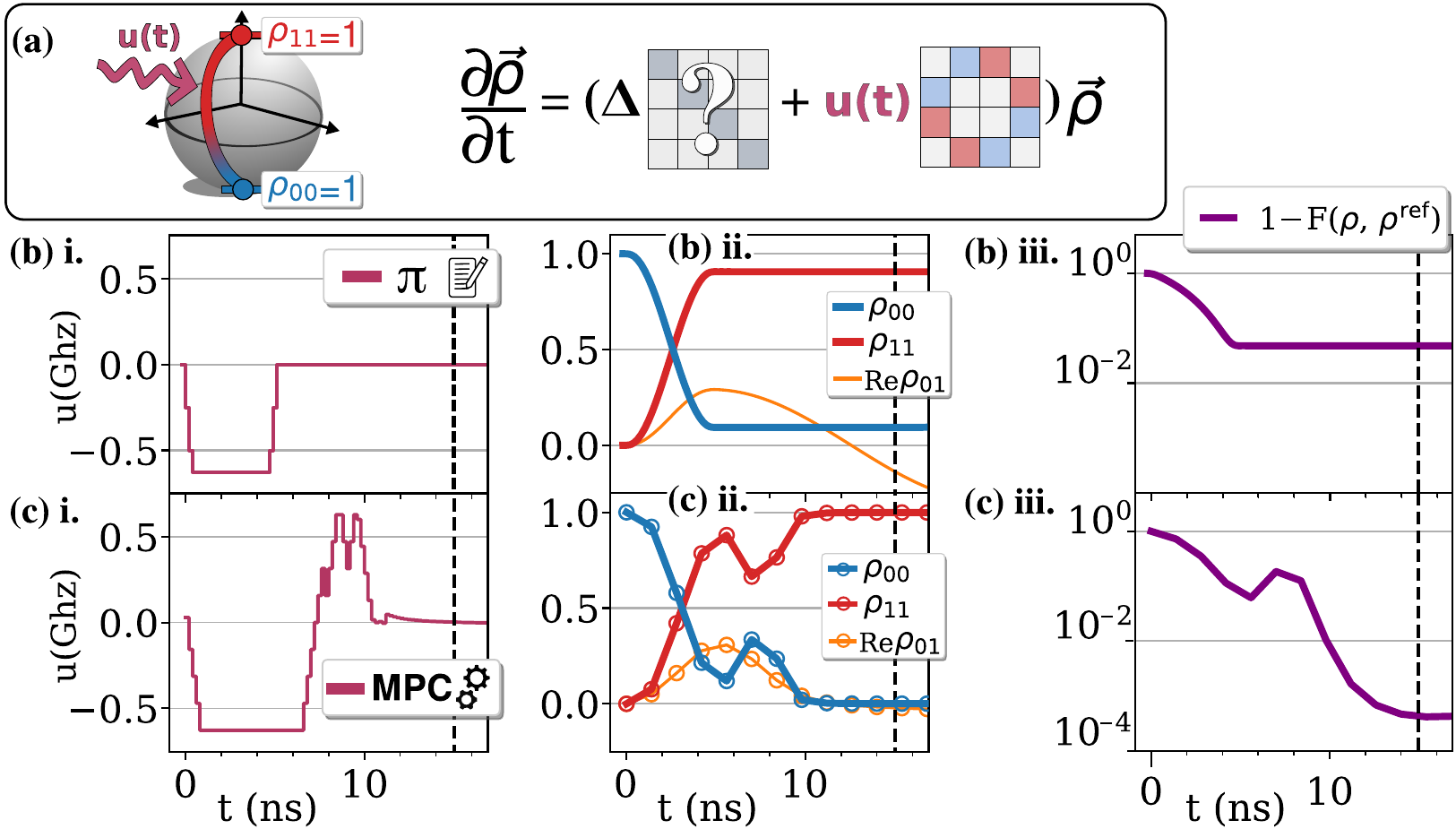}
    \caption{\textbf{(a)}~A qubit trajectory for the desired state preparation is represented on the Bloch sphere. The goal is to transfer the ground state probability to the excited state, i.e. $\rho_{00} \rightarrow \rho_{11}$. There is a model discrepancy of $\abs{\Delta}=200$~MHz relative to the simulation. \textbf{(b)~i.}~Without a discrepancy, an area-$\pi$ pulse would perform the desired state preparation. Constraints have been enforced on the control's maximum amplitude and first derivative. \textbf{(b)~ii.}~The realized trajectories of the density matrix components under the pulse in \textbf{(b)~i.} are plotted. In \textbf{(b)~iii.} the infidelity is shown to fail to reach the hatched success region due to the unreliable model. \textbf{(c)~i.}~The pulse designed by model predictive control (MPC) is shown. The same constraints used in \textbf{(b)~i.} are imposed here. \textbf{(c)~ii.} Every $1.4$~ns, the density matrix was recorded (open circles) and used to initialize an open-loop optimization over the MPC prediction horizon. \textbf{(c)~iii.}~The operation infidelity is shown to reach the hatched success region due to the robust MPC pulse. The dotted vertical lines throughout \textbf{(b)}-\textbf{(c)} indicates the time used for the infidelity in Appendix~\ref{apdx:fperiod}~Figure~\ref{fig:freq_vs_discrep}.}
    \label{fig:star_plot}
\end{figure}

Our first example is a qubit treated in a rotating frame within the rotating wave approximation. The Hamiltonian is
\begin{equation}
    H(t) = \frac{\Delta}{2} \sigma_z + \frac{u(t)}{2} \sigma_x
\end{equation} 
where $\sigma_z$ and $\sigma_x$ are the usual Pauli matrices~\cite{krantz2019quantum}. The prefactor $\Delta = \omega_Q - \omega_R$ is the discrepancy between the qubit resonance frequency $\omega_Q$ and the chosen rotating frame $\omega_R$. The control $u(t)$ is the envelope of a drive pulse with a carrier frequency of $\omega_R$. Our goal is to apply an area-$\pi$ pulse or swap of the occupation probability of the ground and excited states. To do this, we set a fixed $\rho^\textrm{ref}$ such that $\rho^\textrm{ref}_{11} = 1$ else zero. In all examples, we only enforce on-axis terms (e.g. $\rho_{00}$, $\rho_{11}$) in our cost function by way of the appropriate $\mathbf{Q}$. We set $\mathbf{Q}_f = \mathbf{Q}$, and $\mathbf{R} = 10^{-2}~\1$.

Mischaracterization of the qubit frequency $\omega_Q$ results in an offset rotating frame and a nonzero drift term $\Delta$ in the qubit Hamiltonian (visualized in Figure~\ref{fig:star_plot}(a)). Open-loop controllers rely on highly-accurate models to find good pulse designs.  In contrast, we will demonstrate that MPC does not require optimal open-loop solutions over its prediction horizon to realize good pulse designs; instead, MPC iterations are best understood as planning exercises. In Figure~\ref{fig:star_plot}, MPC is used to design a control pulse swapping the occupation probability of the ground and excited states. The simulated qubit is assumed to be mischaracterized. Specifically, we force MPC to rely on an inaccurate model with $\Delta = 0$ while the simulation actually has been set to $\Delta / 2 \pi = -0.2 / 2 \pi \approx -30$~MHz. For a base comparison, in Figure~\ref{fig:star_plot}(b) we report the effect of an analytic area-$\pi$ pulse. This is the naive design that would be realized by an open-loop controller if the $\Delta=0$ model accurately reflected the simulation; we see that the presence of the nonzero $\Delta$ in the simulation renders this analytic pulse insufficient.

We apply MPC in Figure~\ref{fig:star_plot}(c). We set our prediction horizon to $T=10$~ns. Feedback is provided via the simulated density matrix every $1.4$~ns. In experiment, this feedback would necessitate the use of quantum state tomography~\cite{eisert2020quantum}. We explore the effect of decreasing or increasing the feedback period in Appendix~\ref{apdx:fperiod}. Recall that the MPC controller naturally accommodates constraints like those emerging from hardware limitations. To demonstrate, in Figure~\ref{fig:star_plot} notice that we have set constraints on both the maximum control amplitude ($\abs{u} / 2 \pi = 0.1$) and the allowed initial change in control relative to the previous value ($\Delta u / 2 \pi = 0.04$). We enforce the same constraints for the analytic pulse in Figure~\ref{fig:star_plot}(b). In general, these hard constraints can be turned off or softened (by allowing for some mild infractions of the desired inequalities) to increase the space of feasible controls and state trajectories within MPC. Upon comparing Figure~\ref{fig:star_plot}(c) with Figures~\ref{fig:star_plot}(b), we see how the MPC solution uses feedback to adjust and robustly counteract the model deficiencies observed during its receding-horizon iterations. In the introduction, a conceptual comparison between the forward-looking MPC and backward-looking, model-free algorithms (e.g. Reference~\cite{egger2014adaptive,kelly2014optimal,wittler2021integrated}) was offered. To further elucidate this point, Appendix~\ref{apdx:neldermead} illustrates this comparison by way of the toy qubit system introduced here in this section.

\subsection{Weakly-anharmonic qubit} \label{sec:waqubit}
% ----------------------------------
%
\begin{figure}
    \centering
    \includegraphics[width=\textwidth]{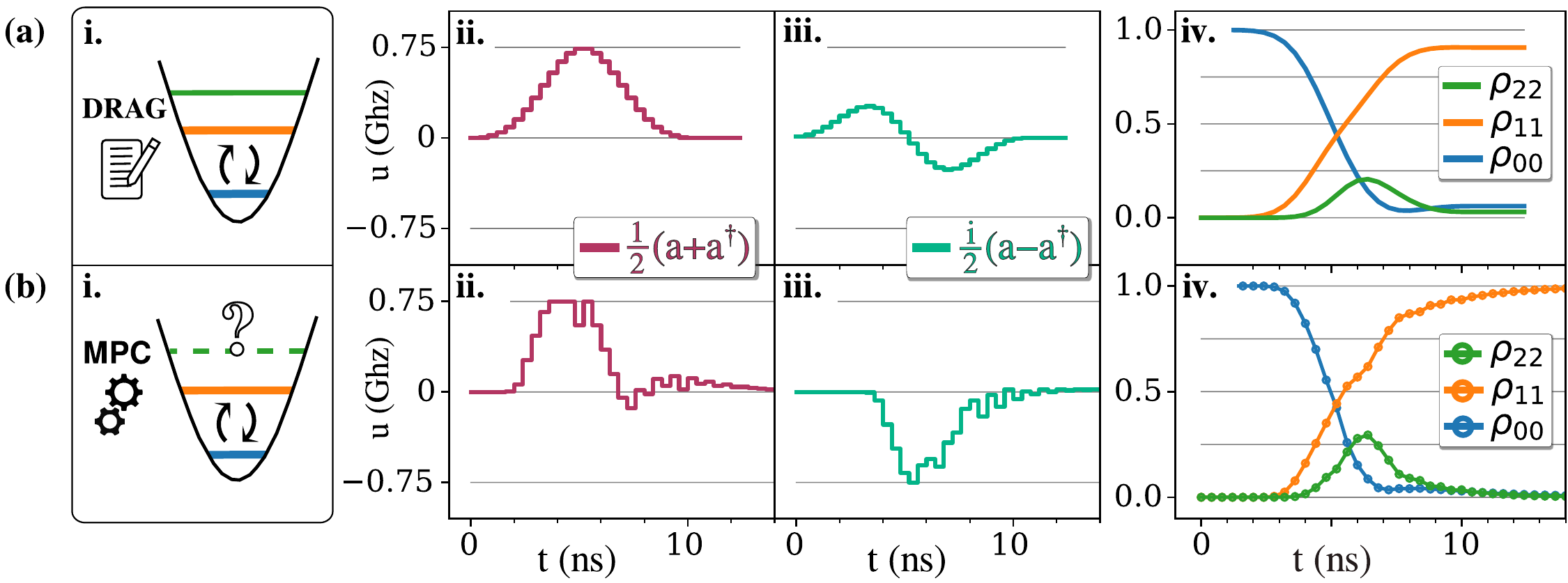}
    \caption{
    \textbf{(a)~i.}~In a weakly-anharmonic oscillator with unequal level spacing quantified by an anharmonicity of $-100$~MHz, the intended outcome is a swap of the ground and excited state probabilities $\rho_{00} \leftrightarrow \rho_{11}$. Here $\rho_{jj}$ is the occupation probability of the $j$-th state, $\ket{j}$.The DRAG procedure is an analytic suppression of leakage out of the qubit subspace defined by levels $\ket{0}$ and $\ket{1}$. It is implemented by a Gaussian $\pi$ pulse on the first control axis in \textbf{(a)~ii.} and a Gaussian derivative on the second control axis in \textbf{(a)~iii.} in order to cancel the spectral overlap of the qubit $\pi$-pulse at the energy splitting of the $\ket{1} \leftrightarrow \ket{2}$ transition. In \textbf{(a)~iv.}, the $\rho_{22}$ population is suppressed and the $\rho_{11}$ is enhanced by the DRAG scheme. In \textbf{(b)~i.}, it is assumed that the anharmonicity is unmodelled and only a qubit model is available. In \textbf{(b)~ii-iii.}, MPC is used to design a pair of robust control pulses by relying on measurement feedback to reduce the leakage from the qubit subspace. \textbf{(b)~iv.} Without modelling the anharmonicity, MPC improves on the state preparation of the analytic DRAG result. Open circles are drawn every $0.4$~ns to indicate the quantum state feedback. Consult Appendix~\ref{apdx:drag} and Figure~\ref{fig:drag_main} for a discussion of additional cases.
    }
    \label{fig:drag_mini}
\end{figure}

A qubit is implemented by restricting a system to a pair of accessible states, for example the ground eigenstate $\ket{0}$ and the first excited eigenstate $\ket{1}$ of a bare system Hamiltonian. If there existed higher level spacing equal to the $\ket{0} \leftrightarrow \ket{1}$ transition, then driving the qubit at the $\ket{0} \leftrightarrow \ket{1}$ frequency would lead to unwanted dynamics involving higher energy eigenstates. For this reason, the energy level spacing of a candidate qubit Hamiltonian must be nonlinear. In transmon qubits, the anharmonicity is the defined to be the difference between the $\ket{1} \leftrightarrow \ket{2}$ level spacing and the $\ket{0} \leftrightarrow \ket{1}$ level spacing of the system's bare Hamiltonian. The limit of infinite absolute anharmonicity results in a perfect qubit; in practice, transmon anharmonicities cannot be made arbitrarily large and are usually between $-100$ to $-300$~MHz~\cite{krantz2019quantum}. Naively, the transmon anharmonicity isolates the qubit transition from exchange with higher energy levels. In practice, fast changes of the drive-pulse envelope widen the spectrum of the qubit control pulse in Fourier space. The wide spectrum can lead to a nontrivial overlap with the frequency of the $\ket{1} \leftrightarrow \ket{2}$ transition and results in an undesired interaction involving the $\ket{2}$ state. The DRAG procedure (Derivative Reduction by Adiabatic Gate) is an analytic suppression of such leakage errors by coordinating control on both the $\sigma_x$ and $\sigma_y$ axes~\cite{motzoi2009simple, krantz2019quantum}. In DRAG, if the first control envelope is fixed to a Gaussian, then the second pulse is proportional to a Gaussian derivative term that eliminates the spectral weight at the $\ket{1} \leftrightarrow \ket{2}$ transition. Numerical optimal control experiments have been shown to reproduce pulse designs similar to the DRAG scheme~\cite{motzoi2009simple,werninghaus2021leakage}.

In Figure~\ref{fig:drag_mini}, we consider a weakly-anharmonic transmon qubit. Our simulation Hamiltonian is defined within the $\ket{0}, \ket{1}, \ket{2}$ space. We use the rotating frame of the $\ket{0} \leftrightarrow \ket{1}$ transition and take the rotating wave approximation to realize
\begin{equation}
    H(t) = \alpha \ket{2}\bra{2} + \frac{u_x(t)}{2}\left(a + a^\dag \right) + \frac{i u_y(t)}{2}\left(a - a^\dag \right)
\end{equation}
where $u_x(t)$ and $u_y(t)$ are the envelopes of pulses driven at the $\ket{0} \leftrightarrow \ket{1}$ transition. In the $\ket{0}, \ket{1}, \ket{2}$ space, the transmon control operations are expressed in terms of truncated raising and lowering operators as $\sigma_x \mapsto (a + a^\dagger) / 2$ and $\sigma_y \mapsto i (a - a^\dagger) / 2$. The transmon anharmonicity is $\alpha=-0.6$ or $\alpha / 2 \pi \approx -100$~MHz.
To demonstrate the robustness of MPC, we suppose that the transmon anharmonicity is unmodelled. That is, our MPC framework is forced to use a model with $\alpha=0$. This is similar to the situation demonstrated in Section~\ref{sec:qubit}, but now there are two control pulses that must operate in tandem. With only a qubit model available to our controller, we rely on MPC to design robust pulses for swapping the ground and excited state probabilities. The prediction horizon is set to $10$ timesteps ($4$~ns) and the feedback timestep is $0.4$~ns. Like in Section~\ref{sec:qubit}, we limit the maximum control amplitude ($\abs{u} \le 0.75$) and constrain the allowed initial change in control relative to its previous value ($\abs{\Delta u} \le 0.2$). Our constraints are chosen so that the MPC pulses are similar to the $10$~ns analytic DRAG pulses shown in Figure~\ref{fig:drag_mini}(a). To implement our DRAG scheme, we set $u_y(t) = 0.6 \dot{u}_x(t) / \abs{\alpha}$. The constant of proportionality is a dimensionless scaling parameter, which we set to the value in $[0,1]$ that maximizes the fidelity of the state preparation~\cite{krantz2019quantum}. In comparing Figure~\ref{fig:drag_mini}(a) to Figure~\ref{fig:drag_mini}(b), we see that MPC is able to coordinate the two control pulses and design a trajectory for the occupation probabilities similar to that of the analytic scheme; significantly, MPC accomplishes this without knowledge of the simulation anharmonicity. In fact, we observe the MPC pulse outperforms our simple version of the analytic DRAG scheme despite this hindrance. In Appendix~\ref{apdx:drag} we add additional context by successfully applying MPC to two more cases: (i) further reducing and (ii) increasing model awareness of the underlying simulation.

\subsection{Crosstalk} \label{sec:crosstalk}
% --------------------
%
\begin{figure}
    \centering
    \includegraphics[width=\textwidth]{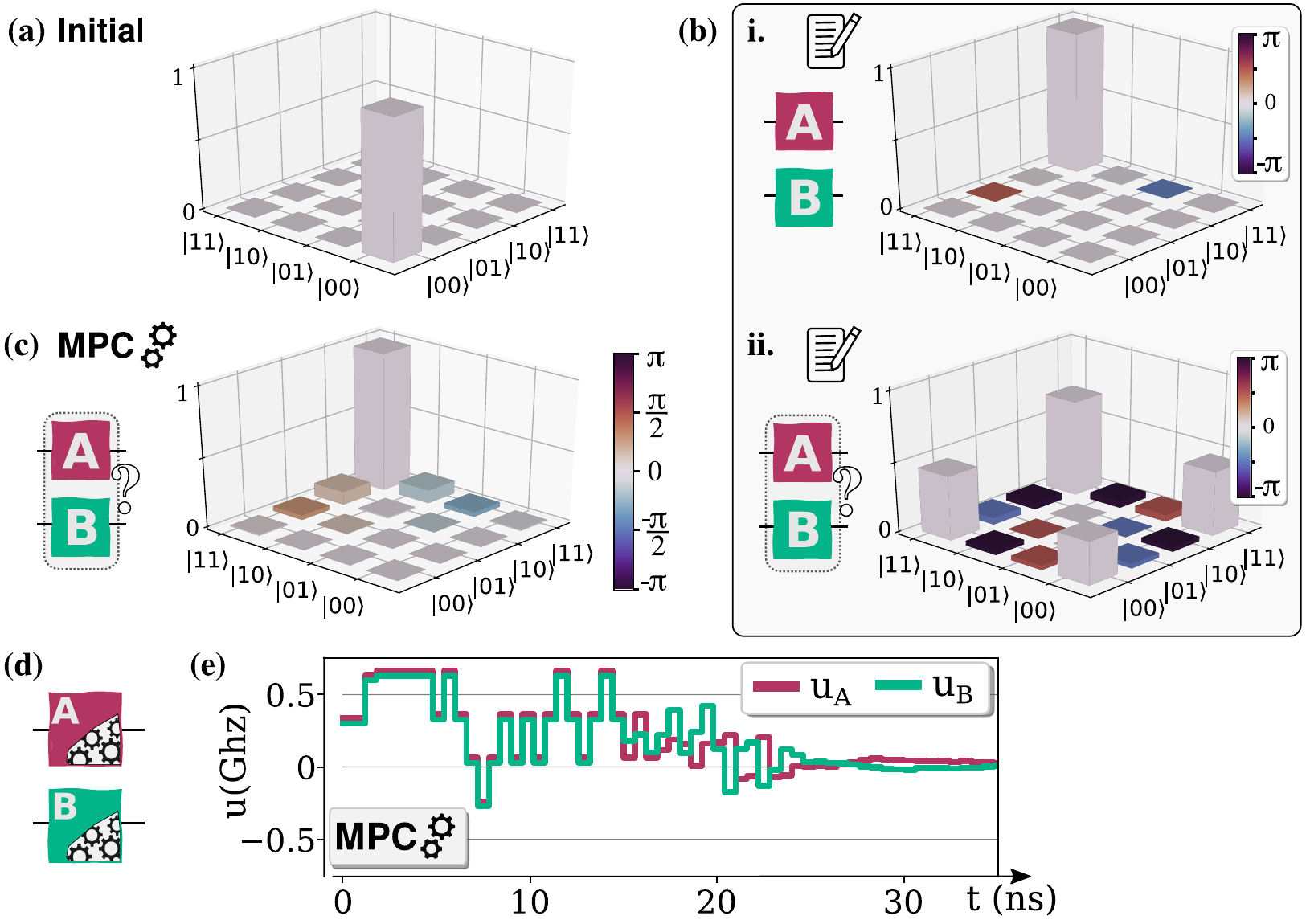}
    \caption{
     A simulation of two qubit state preparations denoted by $\mathbf{A}$ (Hamiltonian: $H_A(t) = u_A(t) \sigma^A_x / 2$) and $\mathbf{B}$ ($H_B(t) = u_B(t) \sigma^B_y / 2$) is modified by a strong crosstalk term proportional to $\sigma^A_z \otimes \sigma^B_z$. \textbf{(a)} The initial density matrix is in the ground state $\rho(0) = \ket{00} \bra{00}$ where $\ket{00} \equiv \ket{0}_A \otimes \ket{0}_B$. The height of the bar indicates the magnitude of that entry in the density matrix, and the color is the phase. The goal is to send the ground state to the excited state for both qubits. This means the target or reference state is set to $\rho^\textrm{ref} = \ket{11}\bra{11}$.
     \textbf{(b)} Suppose the crosstalk term is not present so $u_A$, $u_B$ are analytically two $\pi$-pulses realizing the independent qubit state preparations. \textbf{(b)~i.} In a simulation without crosstalk, the density matrix the coupled system (shown at $25$~ns) is successfully prepared in the reference state. \textbf{(b)~ii.} In a simulation with crosstalk, the same pulses fail to correctly prepare the density matrix of the coupled system (shown at $25$~ns). 
     \textbf{(c)} We force model predictive control (MPC) to use a model with zero crosstalk to design the controls for a simulation where crosstalk is present. The state of the coupled system (shown at $25$~ns) is successfully prepared by MPC. \textbf{(d)} The MPC framework is able to robustly prepare the state using only feedback from the concatenation of the reduced density matrices $\rho_A$, $\rho_B$ in order to overcome the crosstalk modelling discrepancy. \textbf{(e)} The settling time of the successful MPC control is $>20$~ns (shaded region); without crosstalk, the settling time was $<10$~ns.
    }
    \label{fig:crosstalk}
\end{figure}

When calibrating gates and algorithms for quantum processors at scale, additional terms in the subspace dynamics may appear due to unintended crosstalk between otherwise independent parts of the full quantum system~\cite{sarovar2020detecting}. We suggest simultaneously applying MPC to the parts that would, without crosstalk, have known models and independent control objectives. For this purpose, we recall the reduced density matrix $\Tr_E\{\rho_{SE}\} = \rho_{S}$ for a quantum state $\rho \in \mathcal{H}_S \otimes \mathcal{H}_E$ in the joint Hilbert space of a system $S$ and environment $E$. Here, $\Tr_E$ is the partial trace over $\mathcal{H}_E$. In the following discussion, the feedback state supplied to MPC is the concatenation of the reduced density matrices for each of the otherwise independent parts of the full quantum system (i.e. for two parts $A$ and $B$ of a joint system, the feedback state is the concatenation of $\rho_A = \Tr_B\{\rho_{AB}\}$ and $\rho_B = \Tr_A\{\rho_{AB}\}$ as will be described below). As the MPC control horizon recedes, control decisions are made using just the information in these parts to correct for the unintended effects of the crosstalk. At a given time step in the control design, we assume that any observed crosstalk effects were coherent and fixed by the past control decisions. That is, repeat experiments experience the same crosstalk. Corrections based on MPC are made by way of the future evolution. This feature is important because any changes made to past controls modify the form of the past crosstalk in new and uncertain ways.
 
In Figure~\ref{fig:crosstalk}, we use MPC to implement state preparation for two qubits coupled by unintended crosstalk. The qubit spaces are indexed by $A$ and $B$, and the desired state preparations are denoted redundantly as $\textbf{A}$ and $\textbf{B}$. The total system Hamiltonian (each qubit in the rotating frame after the rotating wave approximation) is 
\begin{align} 
    H(t) &= H_{AB} + H_A(t) \otimes \1_B + \1_A \otimes H_B(t) \nonumber \\ \label{eqn:crosstalk}
         &= \frac{\xi}{2} \underbrace{\sigma^A_z \otimes \sigma^B_z}_{\textrm{crosstalk}} + \frac{u_A(t)}{2}~\sigma^A_x \otimes \1_B + \frac{u_B(t)}{2}~\1_A \otimes \sigma^B_y
\end{align}
with $u_A(t)$ and $u_B(t)$ the control envelopes of the on-resonance carrier pulses. Here, we choose $\xi = 0.5$ ($\xi / 2 \pi \approx 80$~MHz). We use separate axes to implement the  $\mathbf{A}$ and  $\mathbf{B}$ operations: respectively, $H_A(t) = u_A(t) \sigma^A_x / 2$ and $H_B(t) = u_B(t) \sigma^B_y / 2$. This allows the trajectories of the reduced states to be somewhat distinct, while simultaneously retaining for each operation the intuition developed during Section~\ref{sec:qubit}.  We assume that the model used by MPC is unaware of the crosstalk in the simulation Hamiltonian, and that the feedback state for MPC is the concatenation of the reduced density matrices $\rho_A$ and $\rho_B$ with otherwise independent objectives $\textbf{A}$ and $\textbf{B}$. The MPC prediction horizon is set to $10$ timesteps ($6$~ns) with a feedback timestep of $0.6$~ns and constraints are added identical to Section~\ref{sec:qubit}.
In Figure~\ref{fig:crosstalk}(a), we show the initial density matrix which is in the ground state of the joint system, $\rho_{AB}(0) = \ket{00} \equiv \ket{0}_A \otimes \ket{0}_B$. Expressed in the joint system, the target or reference state is $\rho^\textrm{ref} = \ket{11} \equiv \ket{1}_A \otimes \ket{1}_B$. In Figures~\ref{fig:crosstalk}(b)-(c) we report the result of different state preparation experiments after $25$~ns. For context, recall that the single qubit state preparations performed in Sections~\ref{sec:qubit}-\ref{sec:waqubit} were realized in $<10$~ns. In Figure~\ref{fig:crosstalk}(b), we apply MPC to design a pair of control pulses for the case of a simulation where no crosstalk is present. In this situation, the model used by MPC matches the simulation. The MPC solution is comparable to two analytic $\pi$ pulses applied to both qubits. This is because the two systems were uncoupled, so the feedback state defined by the concatenation of the reduced density matrices $\rho_A$ and $\rho_B$ provides complete information about the independent system dynamics. Indeed, we see in Figure~\ref{fig:crosstalk}(b)~i. that this pair of $\pi$-pulse proxies prepares the appropriate joint reference state. For comparison, Figure~\ref{fig:crosstalk}(b)~ii. shows the result of using the same pulses to control a simulation where a crosstalk term (Equation~\eqref{eqn:crosstalk}) couples the two systems. Observe that this case (which corresponds to using an analytic or open-loop solution based on an incorrect model) results in a failed state preparation.

Contrast Figure~\ref{fig:crosstalk}(b) with Figure~\ref{fig:crosstalk}(c)--in the latter, we apply MPC to realize our reference state in a simulation with crosstalk using a model that has no crosstalk term. The control pulses designed by MPC are able to use simulation feedback to overcome the model discrepancy from the unknown crosstalk. Figure~\ref{fig:crosstalk}(c) shows that the joint density matrix $\rho_{AB}$ of the coupled system after $25$~ns is successfully prepared in the reference state. Figure~\ref{fig:crosstalk}(d) is a cartoon to emphasize that the control pulses are designed using a feedback state defined by the concatenation of just the reduced density matrices $\rho_A$ and $\rho_B$.
In Figure~\ref{fig:crosstalk}(e) we see the control pulses designed by MPC. Observe that the forward-looking MPC addresses the unknown crosstalk by using feedback to compensate for unplanned dynamics. This results in longer settling times $>20$~ns when compared with the single qubit state preparations of the uncoupled system which were realized in $<10$~ns.

\section{Conclusion}
% ==================
%
Model predictive control (MPC) allows for robust quantum state preparation in the presence of model uncertainty. MPC gains a natural degree of disturbance rejection by relying on feedback to update its receding horizon control plan. In Sections~\ref{sec:qubit}-\ref{sec:crosstalk} we demonstrated this using prototypical superconducting qubit control simulations. In our examples, we showed MPC is able to design successful control pulses when forced to rely on a model that is mismatched from the underlying simulation. Our examples focused on setpoint stabilization (realizing a static reference state), but MPC can be utilized for other control strategies like tracking (following a time-dependent reference trajectory) and path following (staying close to any part of a time-independent reference trajectory at all times)~\cite{rakovic2018handbook}.

MPC is able to integrate state-of-the-art trajectory-based optimization approaches within its receding horizon architecture to improve outcomes and computational tractability at scale. Well-designed optimization constraints can increase robustness and can be used to seek time-optimal controls~\cite{propson2022robust}. Tube MPC can be used in the presence of bounded external disturbances to keep the actual state within an invariant \textit{tube} around the nominal trajectory~\cite{mayne2005robust,lopez2019dynamic,fan2020deep}. A state observer included in trajectory-based optimization can help limit the number of measurements needed to estimate the quantum state~\cite{lee1994extended}. For many-qubit systems, trajectory-based controllers have been developed to assist with the exponential increase of the quantum state space~\cite{abdelhafez2019gradient,shi2019optimized,gunther2021quandary}. These ideas will be incorporated in future work. 

MPC is also simple and flexible enough to complement other existing control strategies.
It is often useful to rely on data-driven models in MPC~\cite{Kaiser2018prsa,baumeister2018deep,bieker2020deep}; for quantum dynamics, one direction is through bilinear dynamic mode decomposition~\cite{goldschmidt2021bilinear}.
MPC and reinforcement learning (RL) have different benefits and costs, and the two can benefit from each other~\cite{gorges2017relations}. MPC can provide the \textit{expert} demonstration needed to initialize the learning process in RL approaches~\cite{phan2021model}. RL approaches are successful at improving quantum optimal control~\cite{niu2019universal,baum2021experimental} but there are significant data costs when training this global model-free approach. The MPC architecture introduced here is robust, interpretable, and requires less data than RL-based strategies. Furthermore, model-based strategies like MPC provide greater potential for extrapolation and generalization during control planning.

\section*{Code Availability}
% =========================
Code supporting the findings of this study is openly available at the following URL:

\url{https://github.com/andgoldschmidt/MPC4quantum}.

\section*{Acknowledgements}
% =========================
%
AG acknowledges support from NSF DMR award \#1747426. The authors acknowledge support from the Army Research Office (ARO~W911NF-17-1-0306) and the National Science Foundation AI Institute in Dynamic Systems (Grant No.~2112085). JLD and AG acknowledge support from the National Nuclear Security Administration Advanced Simulation and Computing Beyond Moore's Law program DP1518100. JLD acknowledges additional support under Laboratory Directed Research and Development Grant 19-ERD-013. This work was partially performed under the auspices of the US Department of Energy by Lawrence Livermore National Laboratory under Contract No. DE-AC52-07NA27344.

% ******
\medskip
\bibliographystyle{plainnat}
\bibliography{library}

\begin{thebibliography}{77}
\providecommand{\natexlab}[1]{#1}
\providecommand{\url}[1]{\texttt{#1}}
\expandafter\ifx\csname urlstyle\endcsname\relax
  \providecommand{\doi}[1]{doi: #1}\else
  \providecommand{\doi}{doi: \begingroup \urlstyle{rm}\Url}\fi

\bibitem[Abdelhafez et~al.(2019)Abdelhafez, Schuster, and
  Koch]{abdelhafez2019gradient}
Mohamed Abdelhafez, David~I Schuster, and Jens Koch.
\newblock Gradient-based optimal control of open quantum systems using quantum
  trajectories and automatic differentiation.
\newblock \emph{Physical Review A}, 99\penalty0 (5):\penalty0 052327, 2019.
\newblock \doi{https://doi.org/10.1103/PhysRevA.99.052327}.

\bibitem[Abraham et~al.(2017)Abraham, de~la Torre, and
  Murphey]{abraham2017model}
Ian Abraham, Gerardo de~la Torre, and Todd Murphey.
\newblock Model-based control using {K}oopman operators.
\newblock In \emph{Robotics: {S}cience and {S}ystems {XIII}}. Robotics: Science
  and Systems Foundation, jul 2017.
\newblock \doi{https://doi.org/10.15607/rss.2017.xiii.052}.

\bibitem[Altafini and Ticozzi(2012)]{altafini2012modelling}
Claudio Altafini and Francesco Ticozzi.
\newblock Modeling and control of quantum systems: {A}n introduction.
\newblock \emph{IEEE Transactions on Automatic Control}, 57\penalty0
  (8):\penalty0 1898--1917, 2012.
\newblock \doi{https://doi.org/10.1109/TAC.2012.2195830}.

\bibitem[Anderson and Moore(2007)]{anderson2007optimal}
Brian~DO Anderson and John~B Moore.
\newblock \emph{Optimal control: {L}inear quadratic methods}.
\newblock Courier Corporation, 2007.

\bibitem[Ball et~al.(2021)Ball, Biercuk, Carvalho, Chen, Hush, De~Castro, Li,
  Liebermann, Slatyer, Edmunds, et~al.]{ball2021software}
Harrison Ball, Michael Biercuk, Andre Carvalho, Jiayin Chen, Michael~Robert
  Hush, Leonardo~A De~Castro, Li~Li, Per~J Liebermann, Harry Slatyer, Claire
  Edmunds, et~al.
\newblock Software tools for quantum control: {I}mproving quantum computer
  performance through noise and error suppression.
\newblock \emph{Quantum Science and Technology}, 2021.
\newblock \doi{https://doi.org/10.1088/2058-9565/abdca6}.

\bibitem[Baum et~al.(2021)Baum, Amico, Howell, Hush, Liuzzi, Mundada, Merkh,
  Carvalho, and Biercuk]{baum2021experimental}
Yuval Baum, Mirko Amico, Sean Howell, Michael Hush, Maggie Liuzzi, Pranav
  Mundada, Thomas Merkh, Andre~R.R. Carvalho, and Michael~J. Biercuk.
\newblock Experimental deep reinforcement learning for error-robust gate-set
  design on a superconducting quantum computer.
\newblock \emph{{PRX} Quantum}, 2\penalty0 (4), nov 2021.
\newblock \doi{https://doi.org/10.1103/prxquantum.2.040324}.

\bibitem[Baumeister et~al.(2018)Baumeister, Brunton, and
  Kutz]{baumeister2018deep}
Thomas Baumeister, Steven~L Brunton, and J~Nathan Kutz.
\newblock Deep learning and model predictive control for self-tuning
  mode-locked lasers.
\newblock \emph{JOSA B}, 35\penalty0 (3):\penalty0 617--626, 2018.
\newblock \doi{https://doi.org/10.1364/JOSAB.35.000617}.

\bibitem[Bieker et~al.(2020)Bieker, Peitz, Brunton, Kutz, and
  Dellnitz]{bieker2020deep}
Katharina Bieker, Sebastian Peitz, Steven~L Brunton, J~Nathan Kutz, and Michael
  Dellnitz.
\newblock Deep model predictive flow control with limited sensor data and
  online learning.
\newblock \emph{Theoretical and Computational Fluid Dynamics}, pages 1--15,
  2020.
\newblock \doi{https://doi.org/10.1007/s00162-020-00520-4}.

\bibitem[Boyd et~al.(2004)Boyd, Boyd, and Vandenberghe]{boyd2004convex}
Stephen Boyd, Stephen~P Boyd, and Lieven Vandenberghe.
\newblock \emph{Convex optimization}.
\newblock Cambridge University Press, 2004.
\newblock \doi{https://doi.org/10.1017/CBO9780511804441}.

\bibitem[Bruder et~al.(2021)Bruder, Fu, and Vasudevan]{bruder2021advantages}
Daniel Bruder, Xun Fu, and Ram Vasudevan.
\newblock Advantages of bilinear {K}oopman realizations for the modeling and
  control of systems with unknown dynamics.
\newblock \emph{IEEE Robotics and Automation Letters}, 6\penalty0 (3):\penalty0
  4369--4376, 2021.
\newblock \doi{https://doi.org/10.1109/LRA.2021.3068117}.

\bibitem[Brunton et~al.(2021)Brunton, Budi{\v{s}}i{\'c}, Kaiser, and
  Kutz]{brunton2021modern}
Steven~L Brunton, Marko Budi{\v{s}}i{\'c}, Eurika Kaiser, and J~Nathan Kutz.
\newblock Modern {K}oopman theory for dynamical systems.
\newblock \emph{arXiv preprint arXiv:2102.12086}, 2021.

\bibitem[{\c{C}}imen(2008)]{ccimen2008state}
Tayfun {\c{C}}imen.
\newblock State-dependent {R}iccati equation ({SDRE}) control: {A} survey.
\newblock \emph{IFAC Proceedings Volumes}, 41\penalty0 (2):\penalty0
  3761--3775, 2008.
\newblock \doi{https://doi.org/10.3182/20080706-5-KR-1001.00635}.

\bibitem[d'Alessandro(2021)]{dalessandro2021introduction}
Domenico d'Alessandro.
\newblock \emph{Introduction to quantum control and dynamics}.
\newblock Chapman and {H}all/{CRC}, 2021.

\bibitem[Diamond and Boyd(2016)]{diamond2016cvxpy}
Steven Diamond and Stephen Boyd.
\newblock {CVXPY}: {A} {P}ython-embedded modeling language for convex
  optimization.
\newblock \emph{Journal of Machine Learning Research}, 17\penalty0
  (1):\penalty0 2909–2913, Jan 2016.
\newblock ISSN 1532-4435.

\bibitem[Egger and Wilhelm(2014)]{egger2014adaptive}
Daniel~J Egger and Frank~K Wilhelm.
\newblock Adaptive hybrid optimal quantum control for imprecisely characterized
  systems.
\newblock \emph{{P}hysical {R}eview {L}etters}, 112\penalty0 (24):\penalty0
  240503, 2014.
\newblock \doi{https://doi.org/10.1103/PhysRevLett.112.240503}.

\bibitem[Eisert et~al.(2020)Eisert, Hangleiter, Walk, Roth, Markham, Parekh,
  Chabaud, and Kashefi]{eisert2020quantum}
Jens Eisert, Dominik Hangleiter, Nathan Walk, Ingo Roth, Damian Markham, Rhea
  Parekh, Ulysse Chabaud, and Elham Kashefi.
\newblock Quantum certification and benchmarking.
\newblock \emph{Nature Reviews Physics}, 2\penalty0 (7):\penalty0 382--390,
  2020.
\newblock \doi{https://doi.org/10.1038/s42254-020-0186-4}.

\bibitem[Eren et~al.(2017)Eren, Prach, Ko{\c{c}}er, Rakovi{\'c}, Kayacan, and
  A{\c{c}}{\i}kme{\c{s}}e]{eren2017model}
Utku Eren, Anna Prach, Ba{\c{s}}aran~Bahad{\i}r Ko{\c{c}}er, Sa{\v{s}}a~V
  Rakovi{\'c}, Erdal Kayacan, and Beh{\c{c}}et A{\c{c}}{\i}kme{\c{s}}e.
\newblock Model predictive control in aerospace systems: {C}urrent state and
  opportunities.
\newblock \emph{Journal of Guidance, Control, and Dynamics}, 40\penalty0
  (7):\penalty0 1541--1566, 2017.
\newblock \doi{https://doi.org/10.2514/1.G002507}.

\bibitem[Falcone et~al.(2007)Falcone, Borrelli, Asgari, Tseng, and
  Hrovat]{falcone2007predictive}
Paolo Falcone, Francesco Borrelli, Jahan Asgari, Hongtei~Eric Tseng, and Davor
  Hrovat.
\newblock Predictive active steering control for autonomous vehicle systems.
\newblock \emph{IEEE Transactions on control systems technology}, 15\penalty0
  (3):\penalty0 566--580, 2007.
\newblock \doi{https://doi.org/10.1109/TCST.2007.894653}.

\bibitem[Fan et~al.(2020)Fan, Agha-mohammadi, and Theodorou]{fan2020deep}
David~D Fan, Ali-akbar Agha-mohammadi, and Evangelos~A Theodorou.
\newblock Deep learning tubes for tube {MPC}.
\newblock In \emph{Robotics: Science and Systems XVI (2020)}, 2020.
\newblock \doi{https://doi.org/10.15607/RSS.2020.XVI.087}.

\bibitem[Folkestad and Burdick(2021)]{folkestad2021koopman}
Carl Folkestad and Joel~W Burdick.
\newblock {K}oopman {NMPC}: {K}oopman-based learning and nonlinear model
  predictive control of control-affine systems.
\newblock In \emph{2021 IEEE International Conference on Robotics and
  Automation (ICRA)}, pages 7350--7356. IEEE, 2021.
\newblock \doi{https://doi.org/10.1109/ICRA48506.2021.9562002}.

\bibitem[Giannakis et~al.(2021)Giannakis, Henriksen, Tropp, and
  Ward]{giannakis2021learning}
Dimitris Giannakis, Amelia Henriksen, Joel~A Tropp, and Rachel Ward.
\newblock Learning to forecast dynamical systems from streaming data.
\newblock \emph{arXiv preprint arXiv:2109.09703}, 2021.

\bibitem[Glaser et~al.(2015)Glaser, Boscain, Calarco, Koch, K{\"o}ckenberger,
  Kosloff, Kuprov, Luy, Schirmer, Schulte-Herbr{\"u}ggen,
  et~al.]{glaser2015training}
Steffen~J Glaser, Ugo Boscain, Tommaso Calarco, Christiane~P Koch, Walter
  K{\"o}ckenberger, Ronnie Kosloff, Ilya Kuprov, Burkhard Luy, Sophie Schirmer,
  Thomas Schulte-Herbr{\"u}ggen, et~al.
\newblock Training {S}chr{\"o}dinger’s cat: {Q}uantum optimal control.
\newblock \emph{The European Physical Journal D}, 69\penalty0 (12):\penalty0
  1--24, 2015.
\newblock \doi{https://doi.org/10.1140/epjd/e2015-60464-1}.

\bibitem[Goerz et~al.(2019)Goerz, Basilewitsch, Gago-Encinas, Krauss, Horn,
  Reich, and Koch]{goerz2019krotov}
Michael Goerz, Daniel Basilewitsch, Fernando Gago-Encinas, Matthias~G Krauss,
  Karl~P Horn, Daniel~M Reich, and Christiane Koch.
\newblock Krotov: {A} {P}ython implementation of krotov's method for quantum
  optimal control.
\newblock \emph{SciPost Physics}, 7\penalty0 (6):\penalty0 080, 2019.
\newblock \doi{https://doi.org/10.21468/SciPostPhys.7.6.080}.

\bibitem[Goerz et~al.(2014)Goerz, Reich, and Koch]{goerz2014optimal}
Michael~H Goerz, Daniel~M Reich, and Christiane~P Koch.
\newblock Optimal control theory for a unitary operation under dissipative
  evolution.
\newblock \emph{New Journal of Physics}, 16\penalty0 (5):\penalty0 055012,
  2014.
\newblock \doi{https://doi.org/10.1088/1367-2630/16/5/055012}.

\bibitem[Goldschmidt et~al.(2021)Goldschmidt, Kaiser, Dubois, Brunton, and
  Kutz]{goldschmidt2021bilinear}
Andy Goldschmidt, Eurika Kaiser, Jonathan~L Dubois, Steven~L Brunton, and
  J~Nathan Kutz.
\newblock Bilinear dynamic mode decomposition for quantum control.
\newblock \emph{New Journal of Physics}, 23\penalty0 (3):\penalty0 033035,
  2021.
\newblock \doi{https://doi.org/10.1088/1367-2630/abe972}.

\bibitem[G{\"o}rges(2017)]{gorges2017relations}
Daniel G{\"o}rges.
\newblock Relations between model predictive control and reinforcement
  learning.
\newblock \emph{IFAC-PapersOnLine}, 50\penalty0 (1):\penalty0 4920--4928, 2017.
\newblock \doi{https://doi.org/10.1016/j.ifacol.2017.08.747}.

\bibitem[Gros et~al.(2020)Gros, Zanon, Quirynen, Bemporad, and
  Diehl]{gros2020from}
Sébastien Gros, Mario Zanon, Rien Quirynen, Alberto Bemporad, and Moritz
  Diehl.
\newblock From linear to nonlinear {MPC}: {B}ridging the gap via the real-time
  iteration.
\newblock \emph{International Journal of Control}, 93\penalty0 (1):\penalty0
  62--80, 2020.
\newblock \doi{https://doi.org/10.1080/00207179.2016.1222553}.

\bibitem[Günther et~al.(2021)Günther, Petersson, and
  DuBois]{gunther2021quandary}
Stefanie Günther, N.~Anders Petersson, and Jonathan~L. DuBois.
\newblock Quandary: {A}n open-source {C++} package for high-performance optimal
  control of open quantum systems.
\newblock In \emph{2021 IEEE/ACM Second International Workshop on Quantum
  Computing Software (QCS)}, pages 88--98, 2021.
\newblock \doi{https://doi.org/10.1109/QCS54837.2021.00014}.

\bibitem[Hincks et~al.(2015)Hincks, Granade, Borneman, and
  Cory]{hincks2015controlling}
IN~Hincks, CE~Granade, Troy~W Borneman, and David~G Cory.
\newblock Controlling quantum devices with nonlinear hardware.
\newblock \emph{Physical Review Applied}, 4\penalty0 (2):\penalty0 024012,
  2015.
\newblock \doi{https://doi.org/10.1103/PhysRevApplied.4.024012}.

\bibitem[Horn and Johnson(1991)]{horn1991topics}
Roger~A. Horn and Charles~R. Johnson.
\newblock \emph{Topics in {M}atrix {A}nalysis}.
\newblock Cambridge University Press, 1991.
\newblock \doi{https://doi.org/10.1017/CBO9780511840371}.

\bibitem[Jackson et~al.(2021)Jackson, Punnoose, Neamati, Tracy, Jitosho, and
  Manchester]{jackson2021altro}
Brian~E Jackson, Tarun Punnoose, Daniel Neamati, Kevin Tracy, Rianna Jitosho,
  and Zachary Manchester.
\newblock {ALTRO-C}: {A} fast solver for conic model-predictive control.
\newblock In \emph{International Conference on Robotics and Automation (ICRA),
  Xi’an, China}, page~8, 2021.
\newblock \doi{https://doi.org/10.1109/ICRA48506.2021.9561438}.

\bibitem[Johansson et~al.(2012)Johansson, Nation, and Nori]{johansson2012qutip}
J~Robert Johansson, Paul~D Nation, and Franco Nori.
\newblock {QuTiP}: {A}n open-source {P}ython framework for the dynamics of open
  quantum systems.
\newblock \emph{Computer Physics Communications}, 183\penalty0 (8):\penalty0
  1760--1772, 2012.
\newblock \doi{https://doi.org/10.1016/j.cpc.2012.02.021}.

\bibitem[Johansson et~al.(2013)Johansson, Nation, and
  Nori]{johansson2013qutip2}
J.R. Johansson, P.D. Nation, and Franco Nori.
\newblock {QuTiP} 2: {A} {P}ython framework for the dynamics of open quantum
  systems.
\newblock \emph{Computer Physics Communications}, 184\penalty0 (4):\penalty0
  1234--1240, apr 2013.
\newblock \doi{https://doi.org/10.1016/j.cpc.2012.11.019}.

\bibitem[Kaiser et~al.(2018)Kaiser, Kutz, and Brunton]{Kaiser2018prsa}
Eurika Kaiser, J~Nathan Kutz, and Steven~L Brunton.
\newblock Sparse identification of nonlinear dynamics for model predictive
  control in the low-data limit.
\newblock \emph{Proceedings of the Royal Society of London A}, 474\penalty0
  (2219), 2018.
\newblock \doi{https://doi.org/10.1098/rspa.2018.0335}.

\bibitem[Kelly et~al.(2014)Kelly, Barends, Campbell, Chen, Chen, Chiaro,
  Dunsworth, Fowler, Hoi, Jeffrey, et~al.]{kelly2014optimal}
Julian Kelly, Rami Barends, Brooks Campbell, Yu~Chen, Zijun Chen, Ben Chiaro,
  Andrew Dunsworth, Austin~G Fowler, I-C Hoi, Evan Jeffrey, et~al.
\newblock Optimal quantum control using randomized benchmarking.
\newblock \emph{{P}hysical {R}eview {L}etters}, 112\penalty0 (24):\penalty0
  240504, 2014.
\newblock \doi{https://doi.org/10.1103/PhysRevLett.112.240504}.

\bibitem[Khaneja et~al.(2005)Khaneja, Reiss, Kehlet, Schulte-Herbr{\"u}ggen,
  and Glaser]{khaneja2005optimal}
Navin Khaneja, Timo Reiss, Cindie Kehlet, Thomas Schulte-Herbr{\"u}ggen, and
  Steffen~J Glaser.
\newblock Optimal control of coupled spin dynamics: {D}esign of {NMR} pulse
  sequences by gradient ascent algorithms.
\newblock \emph{Journal of Magnetic Resonance}, 172\penalty0 (2):\penalty0
  296--305, 2005.
\newblock \doi{https://doi.org/10.1016/j.jmr.2004.11.004}.

\bibitem[Kliesch and Roth(2021)]{kliesch2021theory}
Martin Kliesch and Ingo Roth.
\newblock Theory of quantum system certification.
\newblock \emph{PRX Quantum}, 2\penalty0 (1):\penalty0 010201, 2021.
\newblock \doi{https://doi.org/10.1103/PRXQuantum.2.010201}.

\bibitem[Koopman(1931)]{koopman1931hamiltonian}
B.~O. Koopman.
\newblock {H}amiltonian systems and transformation in {H}ilbert space.
\newblock \emph{Proceedings of the National Academy of Sciences}, 17\penalty0
  (5):\penalty0 315--318, may 1931.
\newblock \doi{https://doi.org/10.1073/pnas.17.5.315}.

\bibitem[Koopman and v.~Neumann(1932)]{koopman1932dynamical}
B.~O. Koopman and J.~v.~Neumann.
\newblock Dynamical systems of continuous spectra.
\newblock \emph{Proceedings of the National Academy of Sciences}, 18\penalty0
  (3):\penalty0 255--263, mar 1932.
\newblock \doi{https://doi.org/10.1073/pnas.18.3.255}.

\bibitem[Korda and Mezi{\'c}(2018)]{korda2018linear}
Milan Korda and Igor Mezi{\'c}.
\newblock Linear predictors for nonlinear dynamical systems: {K}oopman operator
  meets model predictive control.
\newblock \emph{Automatica}, 93:\penalty0 149--160, 2018.
\newblock \doi{https://doi.org/10.1016/j.automatica.2018.03.046}.

\bibitem[Krantz et~al.(2019)Krantz, Kjaergaard, Yan, Orlando, Gustavsson, and
  Oliver]{krantz2019quantum}
Philip Krantz, Morten Kjaergaard, Fei Yan, Terry~P Orlando, Simon Gustavsson,
  and William~D Oliver.
\newblock A quantum engineer's guide to superconducting qubits.
\newblock \emph{Applied Physics Reviews}, 6\penalty0 (2):\penalty0 021318,
  2019.
\newblock \doi{https://doi.org/10.1063/1.5089550}.

\bibitem[Lee and Ricker(1994)]{lee1994extended}
Jay~H Lee and N~Lawrence Ricker.
\newblock Extended {K}alman filter based nonlinear model predictive control.
\newblock \emph{Industrial \& Engineering Chemistry Research}, 33\penalty0
  (6):\penalty0 1530--1541, 1994.

\bibitem[Li et~al.(2022)Li, Ahmed, Saraogi, Lambert, Nori, Pitchford, and
  Shammah]{li2022pulse}
Boxi Li, Shahnawaz Ahmed, Sidhant Saraogi, Neill Lambert, Franco Nori,
  Alexander Pitchford, and Nathan Shammah.
\newblock Pulse-level noisy quantum circuits with qutip.
\newblock \emph{Quantum}, 6:\penalty0 630, 2022.
\newblock \doi{https://doi.org/10.22331/q-2022-01-24-630}.

\bibitem[Lopez et~al.(2019)Lopez, Slotine, and How]{lopez2019dynamic}
Brett~T Lopez, Jean-Jacques~E Slotine, and Jonathan~P How.
\newblock Dynamic tube {MPC} for nonlinear systems.
\newblock In \emph{2019 American Control Conference (ACC)}, pages 1655--1662.
  IEEE, 2019.
\newblock \doi{https://doi.org/10.23919/ACC.2019.8814758}.

\bibitem[Machnes et~al.(2018)Machnes, Ass{\'e}mat, Tannor, and
  Wilhelm]{machnes2018tunable}
Shai Machnes, Elie Ass{\'e}mat, David Tannor, and Frank~K Wilhelm.
\newblock Tunable, flexible, and efficient optimization of control pulses for
  practical qubits.
\newblock \emph{{P}hysical {R}eview {L}etters}, 120\penalty0 (15):\penalty0
  150401, 2018.
\newblock \doi{https://doi.org/10.1103/PhysRevLett.120.150401}.

\bibitem[Magesan and Gambetta(2020)]{magesan2020effective}
Easwar Magesan and Jay~M Gambetta.
\newblock Effective {H}amiltonian models of the cross-resonance gate.
\newblock \emph{Physical Review A}, 101\penalty0 (5):\penalty0 052308, 2020.
\newblock \doi{https://doi.org/10.1103/PhysRevA.101.052308}.

\bibitem[Mayne et~al.(2000)Mayne, Rawlings, Rao, and
  Scokaert]{mayne2000constrained}
David~Q Mayne, James~B Rawlings, Christopher~V Rao, and Pierre~OM Scokaert.
\newblock Constrained model predictive control: {S}tability and optimality.
\newblock \emph{Automatica}, 36\penalty0 (6):\penalty0 789--814, 2000.
\newblock \doi{https://doi.org/10.1016/S0005-1098(99)00214-9}.

\bibitem[Mayne et~al.(2005)Mayne, Seron, and Rakovi{\'c}]{mayne2005robust}
David~Q Mayne, Mar{\'\i}a~M Seron, and SV~Rakovi{\'c}.
\newblock Robust model predictive control of constrained linear systems with
  bounded disturbances.
\newblock \emph{Automatica}, 41\penalty0 (2):\penalty0 219--224, 2005.
\newblock \doi{https://doi.org/10.1016/j.automatica.2004.08.019}.

\bibitem[McKay et~al.(2018)McKay, Alexander, Bello, Biercuk, Bishop, Chen,
  Chow, C{\'o}rcoles, Egger, Filipp, et~al.]{mckay2018qiskit}
David~C McKay, Thomas Alexander, Luciano Bello, Michael~J Biercuk, Lev Bishop,
  Jiayin Chen, Jerry~M Chow, Antonio~D C{\'o}rcoles, Daniel Egger, Stefan
  Filipp, et~al.
\newblock {Qiskit} backend specifications for {OpenQASM} and {OpenPulse}
  experiments.
\newblock \emph{arXiv preprint arXiv:1809.03452}, 2018.

\bibitem[Mezi{\'c}(2005)]{Mezic2005nd}
Igor Mezi{\'c}.
\newblock Spectral properties of dynamical systems, model reduction and
  decompositions.
\newblock \emph{Nonlinear Dynamics}, 41\penalty0 (1-3):\penalty0 309--325,
  2005.
\newblock \doi{https://doi.org/10.1007/s11071-005-2824-x}.

\bibitem[Mezic(2013)]{Mezic2013arfm}
Igor Mezic.
\newblock Analysis of fluid flows via spectral properties of the {K}oopman
  operator.
\newblock \emph{Annual Review of Fluid Mechanics}, 45:\penalty0 357--378, 2013.
\newblock \doi{https://doi.org/10.1146/annurev-fluid-011212-140652}.

\bibitem[Moerland et~al.(2020)Moerland, Broekens, and
  Jonker]{moerland2020model}
Thomas~M Moerland, Joost Broekens, and Catholijn~M Jonker.
\newblock Model-based {R}einforcement {L}earning: {A} survey.
\newblock \emph{arXiv preprint arXiv:2006.16712}, 2020.

\bibitem[Motzoi et~al.(2009)Motzoi, Gambetta, Rebentrost, and
  Wilhelm]{motzoi2009simple}
Felix Motzoi, Jay~M Gambetta, Patrick Rebentrost, and Frank~K Wilhelm.
\newblock Simple pulses for elimination of leakage in weakly nonlinear qubits.
\newblock \emph{{P}hysical {R}eview {L}etters}, 103\penalty0 (11):\penalty0
  110501, 2009.
\newblock \doi{https://doi.org/10.1103/PhysRevLett.103.110501}.

\bibitem[Niu et~al.(2019)Niu, Boixo, Smelyanskiy, and Neven]{niu2019universal}
Murphy~Yuezhen Niu, Sergio Boixo, Vadim~N Smelyanskiy, and Hartmut Neven.
\newblock Universal quantum control through deep reinforcement learning.
\newblock \emph{npj Quantum Information}, 5\penalty0 (1):\penalty0 1--8, 2019.
\newblock \doi{https://doi.org/10.1038/s41534-019-0141-3}.

\bibitem[Nocedal and Wright(2006)]{nocedal2006numerical}
Jorge Nocedal and Stephen Wright.
\newblock \emph{Numerical optimization}.
\newblock Springer Science \& Business Media, 2006.
\newblock \doi{https://doi.org/10.1007/b98874}.

\bibitem[N{\"u}ske et~al.(2021)N{\"u}ske, Peitz, Philipp, Schaller, and
  Worthmann]{nuske2021finite}
Feliks N{\"u}ske, Sebastian Peitz, Friedrich Philipp, Manuel Schaller, and Karl
  Worthmann.
\newblock Finite-data error bounds for {K}oopman-based prediction and control.
\newblock \emph{arXiv preprint arXiv:2108.07102}, 2021.

\bibitem[Peitz and Klus(2019)]{peitz2019koopman}
Sebastian Peitz and Stefan Klus.
\newblock Koopman operator-based model reduction for switched-system control of
  {PDE}s.
\newblock \emph{Automatica}, 106:\penalty0 184--191, 2019.
\newblock \doi{https://doi.org/10.1016/j.automatica.2019.05.016}.

\bibitem[Peitz et~al.(2020)Peitz, Otto, and Rowley]{peitz2020data}
Sebastian Peitz, Samuel~E Otto, and Clarence~W Rowley.
\newblock Data-driven model predictive control using interpolated {K}oopman
  generators.
\newblock \emph{SIAM Journal on Applied Dynamical Systems}, 19\penalty0
  (3):\penalty0 2162--2193, 2020.
\newblock \doi{https://doi.org/10.1137/20M1325678}.

\bibitem[Pendergrass et~al.(2016)Pendergrass, Kutz, and
  Brunton]{pendergrass2016streaming}
Seth~D Pendergrass, J~Nathan Kutz, and Steven~L Brunton.
\newblock Streaming {GPU} singular value and dynamic mode decompositions.
\newblock \emph{arXiv preprint arXiv:1612.07875}, 2016.

\bibitem[Phan and Azad(2021)]{phan2021model}
Minh~Q Phan and Seyed Mahdi~B Azad.
\newblock Model predictive {Q}-learning ({MPQ-L}) for bilinear systems.
\newblock In \emph{Modeling, Simulation and Optimization of Complex Processes
  HPSC 2018}, pages 97--115. Springer, 2021.
\newblock \doi{https://doi.org/10.1007/978-3-030-55240-4_5}.

\bibitem[Propson et~al.(2022)Propson, Jackson, Koch, Manchester, and
  Schuster]{propson2022robust}
Thomas Propson, Brian~E Jackson, Jens Koch, Zachary Manchester, and David~I
  Schuster.
\newblock Robust quantum optimal control with trajectory optimization.
\newblock \emph{Physical Review Applied}, 17\penalty0 (1):\penalty0 014036,
  2022.
\newblock \doi{https://doi.org/10.1103/PhysRevApplied.17.014036}.

\bibitem[Qin and Badgwell(2003)]{qin2003survey}
S~Joe Qin and Thomas~A Badgwell.
\newblock A survey of industrial model predictive control technology.
\newblock \emph{Control engineering practice}, 11\penalty0 (7):\penalty0
  733--764, 2003.
\newblock \doi{https://doi.org/10.1016/S0967-0661(02)00186-7}.

\bibitem[Rakovi{\'c} and Levine(2018)]{rakovic2018handbook}
Sa{\v{s}}a~V Rakovi{\'c} and William~S Levine.
\newblock \emph{Handbook of model predictive control}.
\newblock Springer, 2018.
\newblock \doi{https://doi.org/10.1007/978-3-319-77489-3}.

\bibitem[Sarovar et~al.(2020)Sarovar, Proctor, Rudinger, Young, Nielsen, and
  Blume-Kohout]{sarovar2020detecting}
Mohan Sarovar, Timothy Proctor, Kenneth Rudinger, Kevin Young, Erik Nielsen,
  and Robin Blume-Kohout.
\newblock Detecting crosstalk errors in quantum information processors.
\newblock \emph{Quantum}, 4:\penalty0 321, 2020.
\newblock \doi{https://doi.org/10.22331/q-2020-09-11-321}.

\bibitem[Schaller et~al.(2022)Schaller, Worthmann, Philipp, Peitz, and
  N{\"u}ske]{schaller2022guaranteed}
Manuel Schaller, Karl Worthmann, Friedrich Philipp, Sebastian Peitz, and Feliks
  N{\"u}ske.
\newblock Towards efficient and reliable prediction-based control using {eDMD}.
\newblock \emph{arXiv preprint arXiv:2202.09084}, 2022.

\bibitem[Shi et~al.(2019)Shi, Leung, Gokhale, Rossi, Schuster, Hoffmann, and
  Chong]{shi2019optimized}
Yunong Shi, Nelson Leung, Pranav Gokhale, Zane Rossi, David~I Schuster, Henry
  Hoffmann, and Frederic~T Chong.
\newblock Optimized compilation of aggregated instructions for realistic
  quantum computers.
\newblock In \emph{Proceedings of the Twenty-Fourth International Conference on
  Architectural Support for Programming Languages and Operating Systems}, pages
  1031--1044, 2019.
\newblock \doi{https://doi.org/10.1145/3297858.3304018}.

\bibitem[Silv{\'e}rio et~al.(2022)Silv{\'e}rio, Grijalva, Dalyac, Leclerc,
  Karalekas, Shammah, Beji, Henry, and Henriet]{silverio2022pulser}
Henrique Silv{\'e}rio, Sebasti{\'a}n Grijalva, Constantin Dalyac, Lucas
  Leclerc, Peter~J Karalekas, Nathan Shammah, Mourad Beji, Louis-Paul Henry,
  and Lo{\"\i}c Henriet.
\newblock Pulser: An open-source package for the design of pulse sequences in
  programmable neutral-atom arrays.
\newblock \emph{Quantum}, 6:\penalty0 629, 2022.
\newblock \doi{https://doi.org/10.22331/q-2022-01-24-629}.

\bibitem[Stellato et~al.(2020)Stellato, Banjac, Goulart, Bemporad, and
  Boyd]{stellato2021osqp}
B.~Stellato, G.~Banjac, P.~Goulart, A.~Bemporad, and S.~Boyd.
\newblock {OSQP}: {A}n operator splitting solver for quadratic programs.
\newblock \emph{Mathematical Programming Computation}, 12\penalty0
  (4):\penalty0 637--672, 2020.
\newblock \doi{https://doi.org/10.1007/s12532-020-00179-2}.

\bibitem[Virtanen et~al.(2020)Virtanen, Gommers, Oliphant, Haberland, Reddy,
  Cournapeau, Burovski, Peterson, Weckesser, Bright, {van der Walt}, Brett,
  Wilson, Millman, Mayorov, Nelson, Jones, Kern, Larson, Carey, Polat, Feng,
  Moore, {VanderPlas}, Laxalde, Perktold, Cimrman, Henriksen, Quintero, Harris,
  Archibald, Ribeiro, Pedregosa, {van Mulbregt}, and {SciPy 1.0
  Contributors}]{virtanen2020scipy}
Pauli Virtanen, Ralf Gommers, Travis~E. Oliphant, Matt Haberland, Tyler Reddy,
  David Cournapeau, Evgeni Burovski, Pearu Peterson, Warren Weckesser, Jonathan
  Bright, St{\'e}fan~J. {van der Walt}, Matthew Brett, Joshua Wilson, K.~Jarrod
  Millman, Nikolay Mayorov, Andrew R.~J. Nelson, Eric Jones, Robert Kern, Eric
  Larson, C~J Carey, {\.I}lhan Polat, Yu~Feng, Eric~W. Moore, Jake
  {VanderPlas}, Denis Laxalde, Josef Perktold, Robert Cimrman, Ian Henriksen,
  E.~A. Quintero, Charles~R. Harris, Anne~M. Archibald, Ant{\^o}nio~H. Ribeiro,
  Fabian Pedregosa, Paul {van Mulbregt}, and {SciPy 1.0 Contributors}.
\newblock {{SciPy} 1.0: Fundamental Algorithms for Scientific Computing in
  Python}.
\newblock \emph{Nature Methods}, 17:\penalty0 261--272, 2020.
\newblock \doi{https://doi.org/10.1038/s41592-019-0686-2}.

\bibitem[von Neumann et~al.(1932)von Neumann, Beyer, and
  Wheeler]{vonneumann1932mathematical}
John von Neumann, Robert~T. Beyer, and Nicholas~A. Wheeler.
\newblock \emph{Mathematical foundations of quantum mechanics}.
\newblock Princeton University Press, 2018 edition, 1932.
\newblock ISBN 9780691178561.
\newblock \doi{https://doi.org/10.1515/9781400889921}.

\bibitem[Wang and Boyd(2009)]{wang2009fast}
Yang Wang and Stephen Boyd.
\newblock Fast model predictive control using online optimization.
\newblock \emph{IEEE Transactions on control systems technology}, 18\penalty0
  (2):\penalty0 267--278, 2009.
\newblock \doi{https://doi.org/10.1109/TCST.2009.2017934}.

\bibitem[Watter et~al.(2015)Watter, Springenberg, Boedecker, and
  Riedmiller]{watter2015embed}
Manuel Watter, Jost Springenberg, Joschka Boedecker, and Martin Riedmiller.
\newblock {E}mbed to {C}ontrol: {A} locally linear latent dynamics model for
  control from raw images.
\newblock \emph{Advances in {N}eural {I}nformation {P}rocessing {S}ystems}, 28,
  2015.

\bibitem[Werninghaus et~al.(2021)Werninghaus, Egger, Roy, Machnes, Wilhelm, and
  Filipp]{werninghaus2021leakage}
Max Werninghaus, Daniel~J Egger, Federico Roy, Shai Machnes, Frank~K Wilhelm,
  and Stefan Filipp.
\newblock Leakage reduction in fast superconducting qubit gates via optimal
  control.
\newblock \emph{npj Quantum Information}, 7\penalty0 (1):\penalty0 1--6, 2021.
\newblock \doi{https://doi.org/10.1038/s41534-020-00346-2}.

\bibitem[Wittler et~al.(2021)Wittler, Roy, Pack, Werninghaus, Roy, Egger,
  Filipp, Wilhelm, and Machnes]{wittler2021integrated}
Nicolas Wittler, Federico Roy, Kevin Pack, Max Werninghaus, Anurag~Saha Roy,
  Daniel~J Egger, Stefan Filipp, Frank~K Wilhelm, and Shai Machnes.
\newblock Integrated tool set for control, calibration, and characterization of
  quantum devices applied to superconducting qubits.
\newblock \emph{Physical Review Applied}, 15\penalty0 (3):\penalty0 034080,
  2021.
\newblock \doi{https://doi.org/10.1103/PhysRevApplied.15.034080}.

\bibitem[Wu et~al.(2020)Wu, Tomarken, Petersson, Martinez, Rosen, and
  DuBois]{wu2020high}
Xian Wu, SL~Tomarken, N~Anders Petersson, LA~Martinez, Yaniv~J Rosen, and
  Jonathan~L DuBois.
\newblock High-fidelity software-defined quantum logic on a superconducting
  qudit.
\newblock \emph{{P}hysical {R}eview {L}etters}, 125\penalty0 (17):\penalty0
  170502, 2020.
\newblock \doi{https://doi.org/10.1103/PhysRevLett.125.170502}.

\bibitem[Zhang et~al.(2019)Zhang, Rowley, Deem, and
  Cattafesta]{zhang2019online}
Hao Zhang, Clarence~W Rowley, Eric~A Deem, and Louis~N Cattafesta.
\newblock Online dynamic mode decomposition for time-varying systems.
\newblock \emph{SIAM Journal on Applied Dynamical Systems}, 18\penalty0
  (3):\penalty0 1586--1609, 2019.
\newblock \doi{https://doi.org/10.1137/18M1192329}.

\bibitem[Zhang et~al.(2016)Zhang, Kahn, Levine, and Abbeel]{zhang2016learning}
Tianhao Zhang, Gregory Kahn, Sergey Levine, and Pieter Abbeel.
\newblock Learning deep control policies for autonomous aerial vehicles with
  {MPC}-guided policy search.
\newblock In \emph{2016 IEEE international conference on robotics and
  automation (ICRA)}, pages 528--535. IEEE, 2016.
\newblock \doi{https://doi.org/10.1109/ICRA.2016.7487175}.

\end{thebibliography}
% =========

\appendix
\numberwithin{figure}{section}
\numberwithin{equation}{section}

\section{Vectorization} \label{apdx:vec}
% =====================
The vectorization of a matrix is a linear transformation that converts a matrix into a column vector, \mbox{$\textrm{vec}:R^{m \times n} \rightarrow R^{mn}$}. Vectorization is defined to be column-major or row-major, depending on the stacking priority imposed. For example, Python uses row-major vectorization in NumPy's ndarray.flatten: each row of an input array is sequentially concatenated to form one long row. In Dirac notation, vectorization is the isomorphism for when a bra is flipped to a ket, $\ket{j} \bra{k} \cong \ket{j} \otimes \ket{k}$. 
We restate a well-known property of vectorization (c.f.~Lemma~4.3.2~in~\cite{horn1991topics}) here:
\begin{lemma} \label{note:vec}
    For any linear transformation $L: R^{m \times n} \rightarrow R^{p \times q}$ there exists a unique matrix $K(L) \in R^{mn \times pq}$ such that $\ovec{L(B)} = K(L) \ovec{B}$ for any $B \in R^{m \times n}$. 
\end{lemma}
In particular, for the case of left and right matrix multiplication under row-major vectorization we have
\begin{equation} \label{eqn:multiplyvec}
    \ovec{ABC} = \left(A \otimes C\T \right) \ovec{B}
\end{equation}

\section{Trajectory-based optimization} \label{apdx:opnorm}
% =====================================
In Equation~\eqref{eqn:quadraticcost} we perform trajectory-based optimization for state preparation via a cost  $\norm{\mathbf{x} - \mathbf{x}^\textrm{ref}}_2$ defined in terms of the $\ell_2$-norm. The state is $\mathbf{x} \equiv \ovec{\rho}$ as discussed in Section~\ref{sec:quantumdynamics} and Appendix~\ref{apdx:vec}. Commensurate with this trajectory-based optimization, we noted in Section~\ref{sec:quantumdynamics} that state preparation is often described by quantum information theorists in terms of the state fidelity $F(\rho, \rho^\textrm{ref})=\Tr\{\sqrt{\sqrt{\rho} \rho^\textrm{ref} \sqrt{\rho}}\}^2$. Fidelity approximates (by the Fuchs-van~de~Graaf inequalities) the distance measures of quantum states given by the Schatten $p$-norms, $\norm{\rho - \rho^\textrm{ref}}_p = \norm{\sigma(\rho - \rho^\textrm{ref})}_p = (\sum_j \sigma_j^p)^{1/p}$ where $\sigma(\cdot)$ returns the vector of singular values $\sigma_j$ of a given matrix~\cite{kliesch2021theory}. If both $\rho$ and $\rho^\textrm{ref}$ are pure states, then these two perspectives are equivalent, and
\begin{equation}
     \frac{1}{2} \norm{\mathbf{x} - \mathbf{x}^\textrm{ref}}^2_2 =  \frac{1}{2} \norm{\rho - \rho^\textrm{ref}}^2_2 = 1 - F(\rho, \rho^\textrm{ref}).
\end{equation}

\section{Gate synthesis} \label{apdx:gates}
% ======================
%
In this paper, we have focused on quantum state preparation to illustrate the application of MPC for quantum dynamics. In the case of quantum control for the purposes of quantum computation, the more desirable comparison is to quantum gate synthesis. It is known that quantum state preparation can be thought of as a subroutine for gate synthesis. Indeed, gate synthesis can also be defined as the realization of a reference state under the action of control, as we have done for quantum state preparation.  In gate synthesis, the reference state is no longer a density matrix, but is instead a unitary quantum process. For a closed quantum system, a unitary quantum process is equivalent to time evolution, $\rho(t) = U(t) \rho(0) U^\dag(t)$. 
%More generally, for open quantum systems a general quantum process (in the sense of a completely positive map) can be written $\mathcal{U}(\rho) = \sum_k E_k \rho E_k^\dag$ where $\{E_k\}$ are a set of Krauss operators with $\sum_k E_k^\dag E_k = \1$.
From this, observe that the reference state which is prepared by a fixed gate depends on the initial state $\rho(0)$. This means that the synthesized control pulses for a gate must simultaneously realize the state-dependent reference process for all initial states. It has been shown that for gate synthesis, it is sufficient to consider the simultaneous state preparation of just three specific initial states~\cite{goerz2014optimal}. Therefore, to realize gate synthesis, we can directly apply our MPC algorithm to these concatenated states under dynamics given by the appropriate direct sum of the individual models.
Alternatively, consider the analogy that connects the density matrix associated with a pure state to the quantum process associated with a unitary matrix. First, vectorize the unitary matrix (assume row-major vectorization), $\ovec{U} \equiv \sket{U}$. Now the equivalence
\begin{align}
    \rho \equiv \ket{\psi} \bra{\psi} &\leftrightarrow P \equiv \sket{U} \sbra{U} \\
    \dot{\rho} = - i [H, \rho] & \leftrightarrow \dot{P} = - i [H \otimes \1, P]
\end{align}
can be used to directly map our previous work to the new problem. For example, the reference is now a process matrix, $P^\textrm{ref}$. Note that there are different fidelity measures to score the success of quantum gate synthesis, so the $\ell_2$-norm used in the trajectory-based optimization must be interpreted accordingly~\cite{kliesch2021theory}.

\section{Comparison with model-free descent} \label{apdx:neldermead}
% ==========================================
%
\begin{figure}
    \centering
    \includegraphics[width=\textwidth]{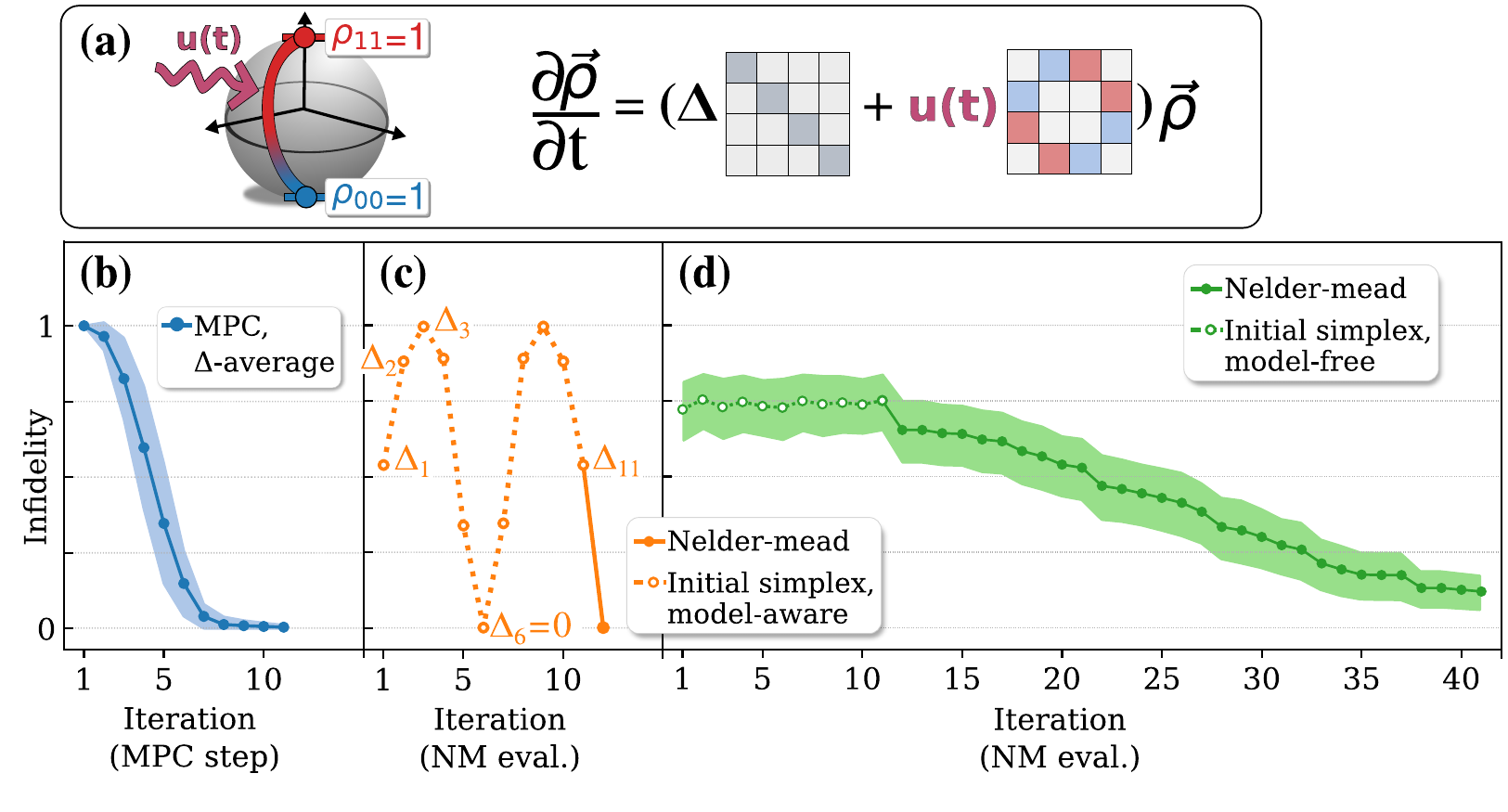}
    \caption{\textbf{(a)} The system from Figure~\ref{fig:star_plot} is repeated here, with $1$~ns timesteps yielding $10$~ns control pulses (thus, $10$ free parameters). The true model is $\Delta = 0$, while a uniform sample of $11$ discrepancies is considered: $\Delta \in [-0.36, 0.36]$. \textbf{(b)} MPC is applied to synthesize a robust $10$~ns control pulse for each of the $11$ models. \textbf{(c)} Nelder-Mead is used for descent in the $10$-dimensional parameter space of the controls. The $11$ coordinates for the initial Nelder-Mead simplex are the open-loop optimal controls determined from the models specified by the $11$ values for $\Delta$ (thus, the simplex is model-aware). Notice the range of infidelities depending on how well the model reflects the simulation. \textbf{(d)} Nelder-Mead is applied using a random initial simplex (thus, the simplex is model-free). Each iteration is an evaluation of the final infidelity of a $10$~ns control pulse.}
    \label{fig:nm_comparison}
\end{figure}

In this appendix, we compare MPC for pulse synthesis to model-free descent in the parameter space of the control pulse. This latter perspective is a simplification of the approach suggested in, e.g. References~\cite{egger2014adaptive,kelly2014optimal,wittler2021integrated}. We compare the two approaches by tracking the iterations of quantum state tomography (QST) required to implement a control pulse that satisfies the state preparation objective from Section~\ref{sec:qubit} (Figure~\ref{fig:star_plot}). In MPC, each QST instance is used to evaluate the current state for the purpose of determining the forward-looking open-loop control optimization. Each MPC step requires one QST iteration to assess the current state. In model-free descent, QST is used to determine the final pulse infidelity. Each QST iteration is an evaluation of this infidelity, which occurs after the application of a complete control sequence. It should be noted that one advantage of model-free descent, as advocated in~\cite{egger2014adaptive,kelly2014optimal,wittler2021integrated}, is the use of methods like randomized benchmarking in lieu of QST~\cite{eisert2020quantum} for the purpose of evaluating a proxy for the direct infidelity.

We use the Nelder-Mead algorithm~\cite{egger2014adaptive} to calibrate the control pulses via model free descent. The Nelder-Mead algorithm starts with a simplex of one dimension larger than the number of parameters, and chooses new search directions based on the objective values of this simplex. In this appendix, we slightly modify the example from Section~\ref{sec:qubit}: we take $1$~ns timesteps and consider a total control window of $10$~ns, meaning that we have $10$ independent control parameters determining our desired pulse. Notice that this means a direct application of the Nelder-Mead algorithm will require an $11$-point simplex. We allow for discrepancies, $\Delta \in [-0.36, 0.36]$~GHz, with the true model given by $\Delta=0$ (for more on these choices, see Figure~\ref{fig:freq_vs_discrep} in Appendix~\ref{apdx:fperiod}). In particular, we consider $11$ uniformly-sampled $\Delta_1, \Delta_2, \dots, \Delta_{11}$. In Figure~\ref{fig:nm_comparison}~(b) we show the infidelity that results from applying MPC with horizon $5$~ns from step $1$ to step $10$ of the pulse synthesis. The reported infidelity has been averaged over the $11$ uniformly-sampled values $\Delta$, with error given by the standard deviation. For the full range of models, MPC is able to prepare the state after $10$ iterations of QST (equivalently, $10$ steps of the MPC algorithm). In Figure~\ref{fig:nm_comparison}~(c) and (d) we instead consider model-free descent for two different initial simplexes. The first, in Figure~\ref{fig:nm_comparison}~(c), we refer to as model-aware. We take as the coordinates of our simplex the $10$~ns pulses determined by solving the open-loop optimal control problem for each $\Delta_1, \Delta_2, \dots, \Delta_{11}$. Notice that the infidelity of the open-loop control varies depending on how closely the model $\Delta$ reflects the true $\Delta = 0$ of the simulation. We call this approach model-aware because we are in effect reducing the model-free calibration to a trial-and-error search over the discrepancy value. Again, each iteration is an evaluation of the infidelity and therefore treated as a QST instance equivalent to an MPC step: just setting up the simplex demands $11$ QST iterations. In Figure~\ref{fig:nm_comparison}~(d), we seed the construction of the initial simplex using a single random $10$~ns pulse within the allowed range of control amplitudes. The remaining coordinates of the simplex are chosen by separately incrementing each component by the maximum $\Delta u$ (a default approach for implementations of Nelder-Mead like in SciPy~\cite{virtanen2020scipy}). In the example in this appendix, we have set the control saturation at $\abs{u}/ 2 \pi = 0.1$ and $\abs{\Delta u} / 2 \pi = 0.05$, to match the example in the main text. A total of $10$ random initial simplexes are considered, providing the mean and variance of the sample mean reported in Figure~\ref{fig:nm_comparison}~(d).

Observe that in the fully model-free application of the Nelder-Mead descent seen in Figure~\ref{fig:nm_comparison}~(d), relatively many iterations were necessary to successfully search the parameter space. Indeed, if we extrapolate to a case where there are more than $10$ free parameters, even the data required for the initial simplex becomes costly--this without even beginning the descent. We see in Figure~\ref{fig:nm_comparison}~(b) that the alternative MPC perspective solved the quantum problem as if it was an \textit{online} optimization. The forward-looking MPC was able to trade iterative improvements for the possibility of extra online control time by accepting and proceeding from the current quantum state tomography outcome. We see that the resulting MPC control scheme was robust across a range of modeling inaccuracies, as MPC inherits a natural degree of disturbance rejection due to the feedback. Moreover, MPC can handle the many parameters of multi-input, multi-output systems because it is only solving a forward-looking open-loop optimization at each step.

\section{MPC limits: Feedback period and discrepancy} \label{apdx:fperiod}
% ==================================================
%
\begin{figure}
    \centering
    \includegraphics[width=\textwidth]{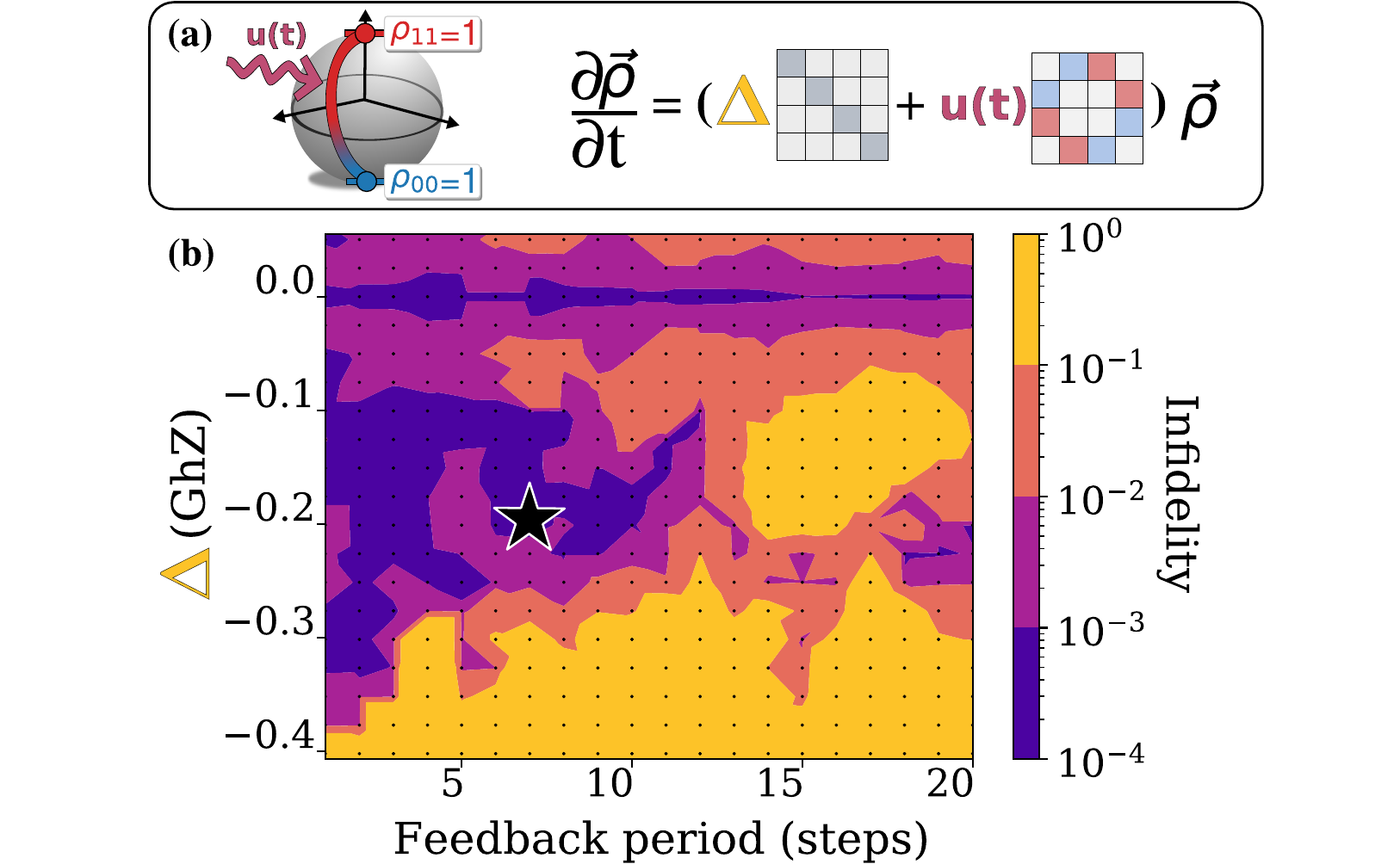}
    \caption{\textbf{(a)} The study from Figure~\ref{fig:star_plot} is repeated here for various $\Delta$ (discrepancies between the model and simulation). \textbf{(b)} Level curves are plotted which show qubit infidelity $F(\rho, \rho^\textrm{ref})$ after $12.5$~ns of the state preparation experiment exemplified by Figure~\ref{fig:star_plot}. The vertical axis, $\Delta$, is the discrepancy between the simulation and model qubit frequencies. The horizontal axis measures the number of model predictive control (MPC) iterations undertaken before receiving the next state feedback from the qubit simulation. For any intermediate iterations, MPC was closed using state feedback from the possibly-inaccurate model predictions.  For example, at $\Delta = 0$~MHz, the model coincided with the simulation so an optimal infidelity (with respect to our numerics) was obtained for all feedback periods. Each dot represents one MPC computation using full sequential quadratic programming (SQP); the coloring was performed via linear interpolation between dots. The results in Figure~\ref{fig:star_plot}(c) were obtained from the parameters labelled by the star.}
    \label{fig:freq_vs_discrep}
\end{figure}

A control strategy based on MPC will eventually break down if the model discrepancy is too great or the feedback is too infrequent. In this appendix, we explore these limits within the context of the example in Section~\ref{sec:qubit}. We consider fixed model discrepancies within MPC. In practice, model discrepancies can be improved by integrating data-driven modelling (e.g., Reference \cite{goldschmidt2021bilinear}) together with MPC: indeed, the error bounds for Koopman models of control systems were recently studied in \cite{nuske2021finite,schaller2022guaranteed}, allowing for a more careful study of data-driven MPC schemes.

In addition to the fixed model discrepancy, the other source of error is the frequency with which the MPC controller incorporates measurement feedback. In applications of MPC for robotics, it is possible that there is a timing mismatch between feedback frequency and MPC runtime. A common challenge is the case where the robot dynamics and corresponding feedback occur faster than the MPC algorithm can return a control decision. Fortunately, the ensemble nature of the quantum experiments used to compute the quantum state tomography limits this concern. On the other hand, in robotics there is also the case where measurements occur slower than the timescales demanded by the model dynamics--a similar situation emerges when trying to limit the number of times the quantum state tomography is performed. In this latter case, the MPC algorithm can be closed using the model of the dynamics until the feedback time is reached, at which point a new quantum state tomography can be performed. Reliance on model predictions can be effective for limiting the number of measurements, but introduces additional dependence on the quality of the model.

In Figure~\ref{fig:freq_vs_discrep}, we looked at the infidelity of the control as a function of the discrepancy $\Delta$ and how often we rely on the model to close the MPC loop. The infidelity of the state preparation was measured after $15$~ns (see Figure~\ref{fig:star_plot}) and is colored by order of magnitude. The horizontal axis reports measurement feedback period in steps (the MPC timestep is $0.2$~ns). This period is the number of MPC iterations for which a model prediction was used to close the loop before a sample from the qubit simulation was used. That is, a long measurement feedback period relies more heavily on a possibly-inaccurate model. Observe that at $\Delta = 0$~MHz in Figure~\ref{fig:freq_vs_discrep} the model prediction coincided with the qubit simulation, so for any measurement feedback period we attain an optimal infidelity (for our numerics).

\section{DRAG} \label{apdx:drag}
% ===========
%
\begin{figure}
    \centering
    \includegraphics[width=\textwidth]{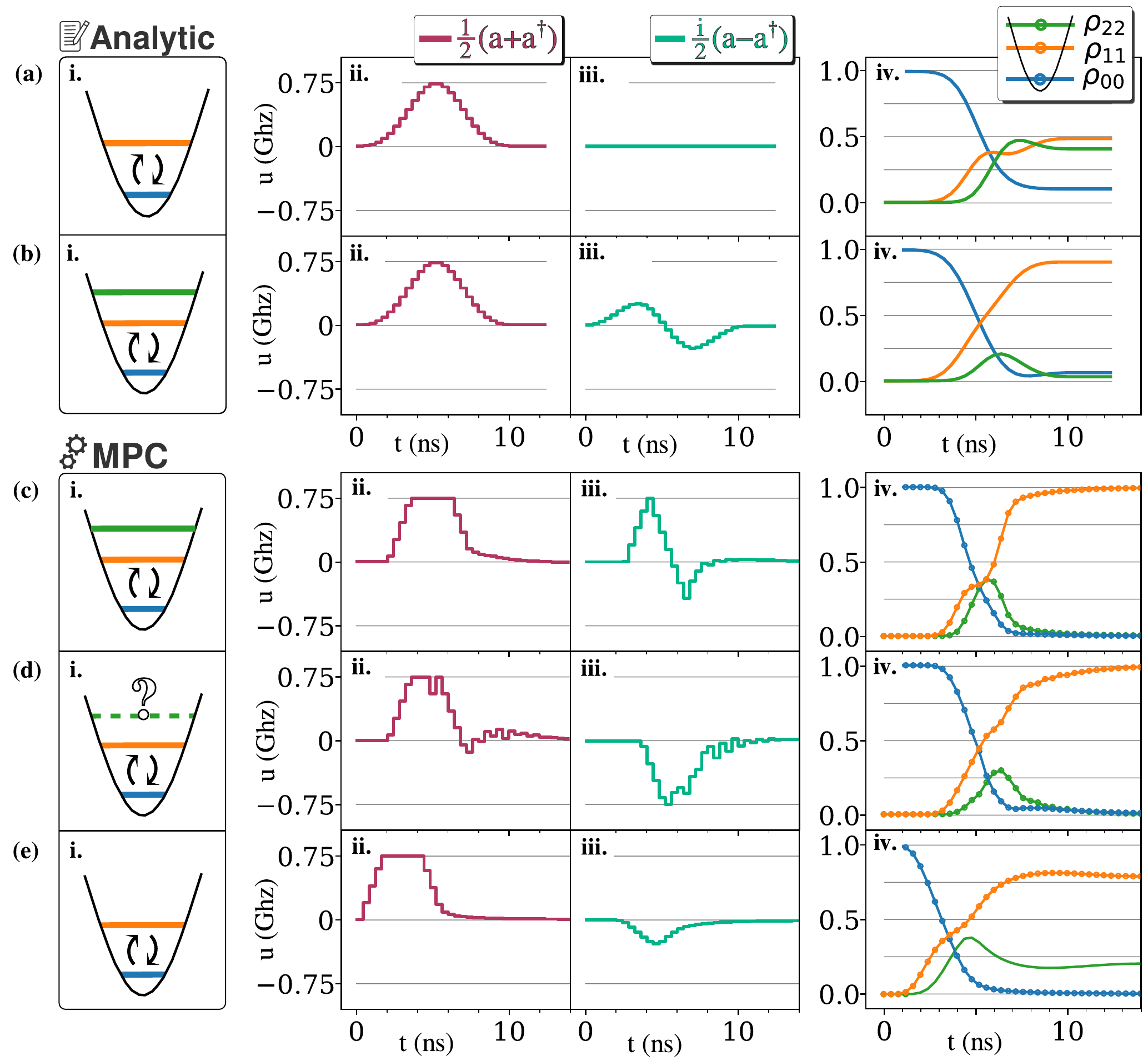}
    \caption{
    This is an expanded version of Figure~\ref{fig:drag_mini} in the main text. In column \textbf{iv.}, recall $\rho_{jj}$ is the occupation probability of the $j$-th state. In \textbf{(a)}, we show the result of failing to include the analytic DRAG correction on the second control term. \textbf{(b)} is the same as Figure~\ref{fig:drag_mini}(a). Parts \textbf{(c)}-\textbf{(e)} rely on MPC under differing assumptions. In the MPC section, column~\textbf{iv.} has open circles on the trajectories to indicate the quantum state feedback used by MPC. In \textbf{(c)}, the model is assumed to match the simulation. Notice that in this case the control design closely resembles the analytic DRAG design in \textbf{(b)}. Part \textbf{(d)} is the same as Figure~\ref{fig:drag_mini}(b). In \textbf{(e)}, the model is restricted to the qubit subspace and MPC is implemented by taking only measurements in this subspace. This is illustrated in \textbf{(e)~iv.} by the absence of open circles over the $\rho_{22}$ trajectory. Even in this limited case, MPC is able to enhance the population of the $\rho_{11}$ state significantly, in contrast to the analytic Gaussian $\pi$-pulse in \textbf{(a)}.
    }
    \label{fig:drag_main}
\end{figure}

Additional exploration of the weakly anharmonic qubit is pursued in this section. Figure~\ref{fig:drag_main} is an expanded version of Figure~\ref{fig:drag_mini} in the main text. New examples appear as new rows, while columns i.-iv. remain the same as the corresponding columns in Figure~\ref{fig:drag_mini} of the main text. An additional analytic example in Figure~\ref{fig:drag_main}(a) shows the result of failing to apply the DRAG correction to the weakly anharmonic qubit. Figure~\ref{fig:drag_main}(b) is the same as Figure~\ref{fig:drag_main}(a) from the main text, showing the result of applying the DRAG scheme. As well, there are two additional examples for MPC. The first is Figure~\ref{fig:drag_main}(c) showing the result of applying MPC in the case where the model used by MPC matches the underlying simulation. A design similar to the DRAG scheme emerges from the numerical optimization. This was also observed in, e.g. References~\cite{motzoi2009simple,werninghaus2021leakage}. Figure~\ref{fig:drag_main}(d) is the same as the main text Figure~\ref{fig:drag_mini}(b) showing MPC in the case where the model does not include the term quantifying the anharmonicity used for the simulation. In Figure~\ref{fig:drag_main}(e), we further reduce the model knowledge by forcing MPC to rely entirely on the qubit subspace. That is, unlike Figure~\ref{fig:drag_main}(d), the feedback state is given only in the reduced qubit subspace and not in the full $\ket{0}, \ket{1}, \ket{2}$ space. Observe in Figure~\ref{fig:drag_main}(e) that a significant enhancement of the $\rho_{11}$ population is still achieved. In such a case as this, where the qubit model is presumed to be the best estimate of the simulation, the resulting analytic pulse is something like the Gaussian $\pi$ pulse shown in Figure~\ref{fig:drag_main}(a). By contrasting Figure~\ref{fig:drag_main}(a) with Figure~\ref{fig:drag_main}(e), we see that MPC provides a robust framework for control design in the presence of model uncertainty.
 
\end{document}